\documentclass[preprint]{aastex}
\usepackage{color}

\def\kms{~{\rm km~s^{-1}}}
\def\cm3{~{\rm cm^{-3}}}
\newcommand{\Mpch}{\mathinner{h^{-1} \mathrm{Mpc}}}
\newcommand{\kpch}{\mathinner{h^{-1} \mathrm{kpc}}}
\newcommand{\Gyr}{\mathinner{\mathrm{Gyr}}}
\newcommand{\rtwo}{\mathinner{r_{200}}}
\newcommand{\mproton}{\mathinner{m_p}}
\newcommand{\pnorm}{\mathinner{\tilde{p}}}
\newcommand{\mJy}{\mathinner{\mathrm{mJy/beam}}}
\newcommand{\Msynctwo}{\mathinner{M_{1.4}^\mathrm{obs}}}
\newcommand{\Msyncthree}{\mathinner{M_{1.4}^\mathrm{w}}}
\newcommand{\Mxtwo}{\mathinner{M_X^\mathrm{obs}}}
\newcommand{\Mxthree}{\mathinner{M_X^\mathrm{w}}}
\newcommand{\Ixbol}{\mathinner{I_{\rm X}}}
\newcommand{\Ssync}{\mathinner{S_{1.4}}}
\newcommand{\rrelic}{\langle \mathinner{R}\rangle}
\newcommand{\Msynctworelic}{\mathinner{M_\mathrm{1.4,relic}^\mathrm{obs}}}
\newcommand{\Msyncthreerelic}{\mathinner{M_\mathrm{1.4,relic}^\mathrm{w}}}
\newcommand{\Mxtworelic}{\mathinner{M_{X,\mathrm{relic}}^\mathrm{obs}}}
\newcommand{\Mxthreerelic}{\mathinner{M_{X,\mathrm{relic}}^\mathrm{w}}}
\newcommand{\Psync}{\mathinner{P_{1.4}}}
\newcommand{\Lxbol}{\mathinner{L_X}}

\newcommand{\lrelic}{\mathinner{l_\mathrm{LLS}}}
\newcommand{\zcl}{\mathinner{z_\mathrm{cl}}}
\newcommand{\microGauss}{\mathinner{\mu \mathrm{G}}}
\newcommand{\fref}[1]{{Figure~\ref{fig:#1}}}
\newcommand{\sref}[1]{{Section~\ref{sec:#1}}}

\begin{document}

\author{Sungwook E. Hong\altaffilmark{1}, Hyesung Kang\altaffilmark{2}, and Dongsu Ryu\altaffilmark{3}\altaffilmark{,4}}
\affil{$^1$School of Physics, Korea Institute for Advanced Study, Seoul 130-722, Korea; swhong@kias.re.kr}
\affil{$^2$Dept. of Earth Sciences, Pusan National University, Busan 609-735, Korea; hskang@pusan.ac.kr}
\affil{$^3$Department of Physics, UNIST, Ulsan 689-798, Korea; ryu@sirius.unist.ac.kr}
\altaffiltext{4}{Author to whom all correspondence should be addressed.}

\title{Radio and X-ray Shocks in Clusters of galaxies}

\begin{abstract}

Radio relics detected in the outskirts of galaxy clusters are thought to trace radio-emitting relativistic
electrons accelerated at cosmological shocks.
{In this study, using the cosmological hydrodynamic simulation data for the large-scale structure formation and 
adopting a diffusive shock acceleration (DSA) model for the production of cosmic-ray (CR) electrons, 
we construct mock radio and X-ray maps of simulated galaxy clusters that are projected in the 
sky plane.
Various properties of shocks and radio relics, including the shock Mach number, radio spectral index 
and luminosity are extracted from the synthetic maps and compared with observations.}
A substantial fraction of radio and X-ray shocks identified in these maps involve multiple shock surfaces 
along line of sights (LoSs), and the morphology of shock distributions in the maps depends on the projection direction.
Among multiple shocks in a given LoS, radio observations tend to pick up stronger shocks with flatter radio spectra,
while X-ray observations preferentially select weaker shocks with larger kinetic energy flux.
{As a result, the shock Mach numbers and locations derived from radio and X-ray observations could differ
from each other in some cases.
We also find that the distributions of the spectral index and radio power of the synthetic radio relics 
are somewhat inconsistent with those of observed real relics; a bit more radio relics have been observed closer to 
the cluster core and with steeper spectral indices.
We suggest the inconsistency could be explained, 
if very weak shocks with $M_s \la 2$ accelerate CR electrons more efficiently, compared to the DSA model adopted here.}
\end{abstract}


\maketitle

\clearpage

\section{Introduction}\label{sec:intro}

Since the discovery of a bow shock in the periphery of the bullet cluster 1E 0657-558 \citep{markevitch2002},
it has been well established that shock waves exist in and around galaxy clusters.
Using cosmological hydrodynamic simulations for the large-scale structure (LSS) formation of the Universe, the origin and nature of shock waves in the intracluster medium (ICM) as well as in the intergalactic medium (IGM) have been extensively studied \citep{miniati2000, ryu2003, pfrommer2006, kang2007, skillman2008, hoeft2008, vazza2009}.
These studies demonstrated that abundant shocks are produced by supersonic flow motions during the process of 
hierarchical clustering of nonlinear structures,
and that they could be classified into two categories, according to their locations relative to the host structures.
\emph{External shocks} are formed at the outermost surfaces surrounding clusters, filaments, and sheets of galaxies
by the accretion of cool ($T \sim 10^3-10^4$ K), tenuous gas in voids onto 
those nonlinear structures.
Since the accretion velocity around clusters can be as high as $v_{\rm acc} \sim$ a few $\times 10^3 \kms$  
and the sound speed of accreting gas is $c_{s}\sim 5-15 \kms$, 
external shocks are strong with Mach number as large as $M_s \sim 10^3$.
On the other hand, \emph{internal shocks} are produced inside the nonlinear structures,
where the gas has been already heated to high temperature by previous episodes of shock passage,
so their Mach number is typically low with $M_s \la 10$.
While most of internal shocks have $M_s \la 3$,
those with $2 \la M_s \la 4$ play the most important role in dissipating the shock kinetic energy 
into heat in the ICM \citep[e.g.,][]{ryu2003,kang2007}.

Internal shocks could be further classified by their origins into a number of types, which in fact may
not be mutually exclusive.
\emph{Turbulent shocks} are induced by turbulent flow motions in the ICM
and they are expected to have very small Mach numbers, $M_s \la 2$ \citep{pjr15}.
Turbulent motions in the ICM could be generated by several different processes: major or minor mergers, AGN jets, 
galactic winds, galaxy wakes, and etc \citep[e.g.,][]{subramanian2006,ryu2008,ryu2012,vazza2012b,brunetti2014}. 
Bow shocks found in the outskirts of merging clusters are commonly referred as \emph{merger shocks}
\citep{roettiger1999,markevitch2007,skillman2013}.
In most cases merger shocks are weak with $M_s \la 3$ \citep[e.g.,][]{gabici2003}.
\emph{Infall shocks} form by infall of the warm-hot intergalactic medium (WHIM; $T \sim 10^5-10^7$ K) 
into the hot ICM along adjacent filaments \citep{hong2014}.
They could have relatively high Mach numbers, reaching up to $M_s \sim 10$,
and thus the ensuing cosmic-ray (CR) acceleration can be more efficient than in other types of shocks.

Including the shock in the bullet cluster, a number of shocks have been found in X-ray observations
\citep[e.g.,][]{russel2010, akamatsu2012, ogrean2013a}.
{In these observations, shocks are detected as sharp discontinuities in the temperature and surface brightness distributions,}
and their physical properties including the sonic Mach number, $M_s$, are estimated from the `deprojected' temperature and density jumps \citep{markevitch2007}.
Most of the shocks detected in X-ray observations are weak with $M_s \sim 1.5-3$.

In addition, shocks in the ICM have been detected by radio observations,
especially as the so-called radio relics \citep[see, e.g.,][for reviews]{feretti2012,bruggen2012}.
Radio relics are usually found in the cluster outskirts around the virial radius, $r_{\rm vir}$.
The radio emission is understood as synchrotron radiation from CR electrons with the 
Lorentz factor of $\gamma_e \sim 10^3-10^5$ that are believed to be accelerated at the shocks
associated with them \citep[e.g.][]{ensslin1998, bagchi06, vanweeren2010}.
High energy nonthermal particles can be produced via diffusive shock acceleration (DSA) at astrophysical shocks,
such as interplanetary and supernova remnant shocks as well as cluster shocks in collisionless tenuous plasma \citep{bell1978, blandford1978, drury1983}.
Moreover, it has been shown that turbulence flow motions in the ICM can produce magnetic fields of up to 
$\sim \microGauss$ level \citep[e.g.,][]{ryu2008}.

{Since the radio-emitting electrons with the Lorentz factor of $\gamma_e \sim 10^3-10^5$}
would not advect nor diffuse in the ICM more than $\sim 100$ kpc 
away from the shock surface before they lose the energy due to
radiative cooling via synchrotron emission and inverse Compton (IC) scattering \citep[e.g.,][]{kang2011},
it is commonly thought that their acceleration sites are likely to be close to where the synchrotron emission is seen.
So the physical properties of shocks in radio relics are inferred from observed quantities such as the injection spectral index at the shock edge, $\alpha_{\rm inj} = ( M_s^2+3)/2(M_s^2-1)$,
and the spatial profile of surface brightness \citep{drury1983, ensslin1998,kang2012}.
Note that the spectral index for the synchrotron spectrum {\it integrated} over the downstream 
behind a steady planar shock is expected to be $\alpha \approx \alpha_{\rm inj}+0.5$ \citep[e.g.][]{kang15b,kangryu2015}.
In most cases, the Mach number of shocks in radio relics was found to be in the range of $M_s \sim 1.5-4.5$ \citep[e.g.,][]{clarke2006, bonafede2009, vanweeren2010, vanweeren2012,stroe14b}.

Pre-acceleration of thermal electrons to suprathermal energies and subsequent injection
into the Fermi first-order process at shock have been one of the outstanding problems in the DSA theory.
It is thought that the particle injection and DSA acceleration might be very inefficient at weak shocks 
($M_s\la 3$) because of small density compression across shocks \citep[e.g.,][]{maldru01}.
Especially, in the so-called ``thermal leakage'' injection model, proton injection is expected to be strongly suppressed at weak shocks;
the relative difference between postshock, proton thermal and flow speeds is greater 
and so it is less likely for postshock protons to recross the shock front at weaker shocks \citep[e.g.,][]{kang2002}.
Although postshock electrons move faster than protons,
they are tied to magnetic field fluctuations more tightly because of
smaller rigidities (i.e. $p_{\rm th,e}=(m_e/m_p)^{1/2} p_{\rm th,p}$).
So it was speculated that the DSA of electrons at weak shocks would not be efficient either, for instance,
not enough to 
explain the observed radio flux of spectacular giant radio relics such as the Sausage relic.

A pre-exiting population of electrons with the Lorentz factor of $\gamma_e \sim 10-100$ was suggested as a possible 
solution to the low electron injection problem at weak cluster shocks \citep{kang2012,pinzke2013}.
Moreover, \citet{kang2014} has suggested that a $\kappa$-like distribution of suprathermal electrons may
exist in high beta ($\beta=P_g/P_B \sim 100$) ICM plasmas just like in solar winds, and facilitate the electron injection at weak shocks.
Recently, using Particle-in-Cell (PIC) simulations of weak shocks in high beta plasmas, \citet{guo14} and \citet{park15} have shown
that some of the incoming electrons gain energy via shock drift acceleration (SDA)
and are reflected specularly toward the upstream region.
Those reflected particles can be scattered back to the shock surface by plasma waves excited in the upstream region, 
and then undergo multiple cycles of SDA, resulting in a power-law type suprathermal population up to $\gamma_e \sim 100$.
These studies suggest a possibility that ``self pre-acceleration'' of thermal electrons to suprathermal energies
via kinetic plasma processes at the shock itself might provide seed electrons 
enough to explain the observed flux level of bright radio relics.

The morphology of radio relics is, in some cases, observed to be elongated and arc-like with a sharp edge 
on one side.
And some of radio relics are found as a pair in the opposite side of clusters.
So they are often interpreted as products of binary mergers \citep[e.g.,][]{ensslin1998, roettiger1999, vanweeren2010, gasperin14}.
Several numerical studies have suggested that merger shocks with sufficient amount of shock kinetic 
energy flux could produce radio relics \citep{nuza2012, vazza2012a, skillman2013}.
In these studies, synthetic radio maps were constructed by 
identifying shocks in simulated clusters and modeling CR electron injection/acceleration and magnetic field strength.
However, there remain a few issues to be resolved,
before the simple picture of merger shocks being the origin of radio relics is accepted.

{The first issue concerns the frequency of observed radio relics. 
Although structure formation simulations have demonstrated that shocks are produced frequently during mergers
and they should last for the cluster dynamical time of $t_{\rm dyn} \sim 1$~Gyr, 
only about $10\%$ of X-ray luminous clusters host some radio relics, putative merger shocks,
and the fraction of merging clusters with giant radio relics is even much lower \citep{feretti2012}.
Recently, it has been suggested that ICM shocks may light up as radio relics only when they
encounter fossil relativistic electrons that are left over from either a
previous episode of shock/turbulence acceleration or a radio jet from AGN \citep{shimwell15,kangryu2015}.
In such a scenario, only a small faction of ICM shocks become radio-emitting structures for a
fraction of the dynamical time ($\la 0.1  t_{\rm dyn}\sim 100$~Myr).
So the rareness of radio relics could be explained.
Of course, this model can be justified only if the injection and acceleration of electrons at weak shocks 
in the ICM is very inefficient, 
so that the re-acceleration of pre-existing CR electrons must be required for the birth of radio relics.}

{The second issue involves the discrepancies in the shock properties inferred from radio and X-ray observations
of a few radio relics.
Among several dozens of observed radio relics,
only a fraction of shocks associated with them also have been detected in X-ray observations 
\citep[][and references therein]{nuza2012, bonafede2012}.
In the case of the Toothbrush relic in 1RXS J0603.3, for exmaple, 
the radio index was measured to be $\alpha_{\rm inj}\approx 0.6-0.7$, indicating a `radio Mach number',
$M_{\rm radio}\approx 3.3-4.6$ \citep{vanweeren2012}.
But the temperature and density discontinuities in X-ray observations suggest an `X-ray Mach number', 
$M_{\rm X} \la 2$, and the position of shock identified in X-ray observations is 
shifted from that in radio observations by $\ga 200$ kpc \citep{akamatsu2013,ogrean2013b}.
In addition, \citet{trasatti15} suggested that for the radio relic in A2256, 
if the observed index $\alpha_{63}^{1360}\approx 0.85$ measured between 63 and 1360~MHz is interpreted 
as the injection index, the radio Mach number can be estimated to be $M_{\rm radio}\approx 2.6$, 
while the temperature jump measured in X-ray observations implies the X-ray Mach number, $M_{\rm X}\sim 1.7$.}
{In the case of the Sausage relic in CIZA J2242.8, \citet {vanweeren2010} used the observed radio spectral index near the edge 
(shock surface), $\alpha_{\rm inj}\approx 0.6$, to obtain a radio-inferred shock Mach number, $M_{\rm radio}\approx 4.6$.
But X-ray-inferred Mach number reported later by Suzaku and Chandra observations 
turned out to be lower with $M_{\rm X-ray}\approx 2.54-3.15$
\citep{akamatsu2013, ogrean14}.
Recently, however, \citet{stroe14a} estimated a steeper value, $\alpha_{\rm inj}\approx 0.77$, 
by performing a spatially-resolved spectral fitting, implying a weaker shock with
$M_{\rm radio}\approx 2.9$ in good agreement with X-ray observations.
}

In \citet[][hereafter, Paper I]{hong2014}, we studied properties of shock waves in the outskirts of simulated galaxy clusters using sets of LSS formation simulations.
{We find that in addition to merger shocks, infall shocks are produced in ICMs by continuous infall of density clumps (i.e. minor mergers)
along filaments of galaxies to hot ICMs.
Unlike weak bow shocks ($M_s \la 2$) driven by major mergers,
infall shocks do not show pairing structures and have higher Mach numbers ($M_s \sim 3 - 10$).
In a few cases (e.g., Coma 1253+275 and NGC 1265), observed radio relics are thought to be associated with infall shocks \citep{brown2011, pfrommer2011, ogrean2013a}.}

{In this paper, we consider a scenario in which suprathermal electrons are injected directly at the shocks
without a help of fossil CR electrons, and accelerated to radio-emitting energies.
Then the frequency of radio relics depends on the shock statistics, 
while the radio luminosity of radio relics is determined by 
the shock kinetic energy flux and the assumed DSA efficiency model \citep[e.g.][]{kang2013}.
In this scenario the rareness of radio relics, compared to the frequency of the ICM shocks, can be
controlled by adjusting the models for the DSA acceleration efficiency and the magnetic field amplification.
For instance, \citet{vazza2012a} showed, using structure formation simulations and
generating mock radio maps of simulated cluster samples, that radio emission
tends to increase toward the cluster periphery and peak around $0.2-0.5r_{\rm vir}$,
mainly because the kinetic energy dissipated
at shocks peaks around $0.2R_{\rm vir}$ and the Mach number of shocks tend to increase toward the cluster outskirts.
Such findings can explain why radio relics are rarely found in the cluster central regions.}

{We study the properties of radio and X-ray shocks
in the simulated cluster sample of Paper I, based on a DSA model in which suprathermal electrons are injected and accelerated
preferentially at shocks with higher Mach numbers.}
We first calculate synchrotron emission at shocks by modeling primary CR electron population based on recent DSA simulations \citep{kang2013} and magnetic field distribution based on a turbulent dynamo model \citep{ryu2008}.
We produce mock radio and X-ray maps by projecting the synchrotron and 
bremsstrahlung emission, respectively.
We then identify ``synthetic radio relics'' in the radio map
and extract their properties, such as radio and X-ray shock Mach numbers, $M_{\rm radio}$ and $M_{\rm X}$, 
and their locations in the map.
{We examine the radio and X-ray properties of relic shocks, and
attempt to understand discrepancies derived from radio and X-ray observations in some radio relics.}

In \sref{numerics}, we present numerical details such as the calculation of synchrotron emission at shock, 
the construction of mock radio and X-ray maps, and the extraction of radio and X-ray shock Mach numbers.
In \sref{projection}, we discuss how the two-dimensional (2D) projection of three-dimensional (3D) shock distributions affects the radio and X-ray observations of ICM shocks.
In \sref{stats}, we describe the properties of shocks in 2D projection
as well as the properties for synthetic radio relics.
Summary follows in \sref{summary}.

\section{Numerical Details}\label{sec:numerics}

\subsection{Clusters and Shocks}\label{sec:numerics:cluster}

The construction of galaxy cluster sample and the identification of shocks are described in detail in Paper I, so here they are summarized only briefly.
The LSS formation simulations adopted a standard $\Lambda$CDM model
with the following parameters:
baryon density fraction $\Omega_\mathrm{BM} = 0.044$, 
dark matter density fraction $\Omega_\mathrm{DM} = 0.236$, 
cosmological constant fraction $\Omega_\Lambda = 0.72$,
Hubble parameter $h \equiv H_0 / (100 \mathrm{km~s^{-1}Mpc^{-1}}) = 0.7$,
rms density fluctuation $\sigma_8 = 0.82$,
and the primordial spectral index $n = 0.96$.
{An updated version of a particle-mesh/Eulerian hydrodynamic 
code described in \cite{ryu1993} was used for the simulations.}
Three sets of simulations were performed:
(1) 16 different realizations of LSS formation in a comoving cubic box of $L=100 \Mpch$ with $1024^3$ uniform grid zones,
(2) 16 different realizations in a comoving cubic box of $L=200 \Mpch$ with $1024^3$ zones,
and (3) one realizations in a comoving cubic box of $L=100 \Mpch$ with $2048^3$ zones.
{The sets of (1) and (2) were performed in an adiabatic (non-radiative) way,
while (3) includes a mild feedback from star formation and cooling/heating processes.
In Paper I, we find that, in all three types of simulations, the statistics of the physical 
properties of clusters and shocks are similar.}

In the simulation data, clusters are identified as regions around local peaks in the spatial distribution of X-ray emissivity.
Around each peak, the X-ray emission-weighted mean temperature,
$T_{\rm X,cl}$, is obtained for the spherical volume within $r\le \rtwo$.
Here, $\rtwo\approx 1.3 r_{\rm vir}$ is the radius within which
the gas overdensity is 200 times the mean gas density of the universe.
Those with $k_B T_{\rm X,cl} \geq 2$ keV for $100 \Mpch$ box simulations and $k_B T_{\rm X,cl} \geq 4$ keV for $200 \Mpch$ box simulations
are selected as synthetic clusters, resulting in a total of 228 clusters from the three
sets of simulations. Here, $k_B$ is the Boltzmann constant

Shocks (actually, shock zones) are identified by a set of criteria given in Paper I, 
and their sonic Mach numbers, $M_s$, are estimated from 
the temperature jump condition,
${T_2}/{T_1} = {(5M_s^2 -1)(M_s^2 + 3)}$ $/({16 M_s^2})$,
where $T$ is the gas temperature.
Hereafter, the subscripts ``1'' and ``2'' indicate the preshock and postshock quantities, respectively.
Only shocks with $M_s \ge 1.5$ are considered.
Note that a shock surface normally consists of many of these shock zones.
The shock speed and shock kinetic energy flux at shock zone are, then, calculated as
$v_1 = M_s (\gamma P_\mathrm{th, 1}/\rho_1)^{1/2}$ and 
$f_\mathrm{kin} = (1/2) \rho_1 v_1^3$, respectively,
where $P_\mathrm{th,1}$ is the preshock thermal pressure
and $\gamma = 5/3$ is the gas adiabatic index.
The energy flux of CR protons can be estimated as
\begin{equation}
f_{{\rm CR},p} = \eta(M_s) \cdot f_\mathrm{kin}=  \eta(M_s) \cdot (1/2) \rho_1 v_1^3,
\label{fcrp}
\end{equation}
where $\eta(M_s)$ is the CR acceleration efficiency via the DSA process.
For the efficiency, the values presented in \citet{kang2013} (also shown in Figure 2 of Paper I) are adopted.

In Paper I, 
we find that the morphology of shock surfaces is quite complex due to the dynamic history of clusters,
and in general a connected shock surface can consist of a number of grid zones of
different types of shocks including merger shocks (see Figure 7 in Paper I).

\subsection{Modeling of Magnetic Field Strength}\label{sec:numerics:magnetic}

{To calculate radio synchrotron emission at cluster shocks,
we need to model the strength of magnetic fields as well as the energy spectrum of CR electrons.
The ICM is observed to be permeated with magnetic fields of a few $\mu$G in the cluster core region
and a few $\times\ 0.1\ \mu$G in the cluster periphery \citep[e.g.][]{bonafede11,feretti2012}.
Suggested ideas for the generation and amplification of magnetic fields in the ICM include 
processes during primordial phase transitions, Biermann battery mechanism, plasma processes at collisionless shocks,  
different types of turbulence dynamo, and 
the ejection of galactic winds and AGN jets \citep[e.g.][]{dolag08,ryu2008,ryu2012,widro12}.}

{Here we adopt the turbulent dynamo model of \citet{ryu2008}, which assumes that
turbulent flow motions are induced via stretching and compression of the vorticity 
generated behind curved surfaces of shocks in clusters, and the magnetic fields are amplified by the turbulent flows.
Then the strength of the resulting magnetic fields can be modeled in terms of the number of local 
eddy turn-over by the following fitting formula: 
\begin{equation}
{B^2 \over 8 \pi \epsilon_{\rm turb}} \equiv \phi(t/t_{\rm eddy})
\approx \left\{ \begin{array}{ll}  0.04 \cdot \exp[(t/t_{\rm eddy}-4)/0.36] & \textrm{ for } t/t_{\rm eddy} < 4 \\
    (0.36/41) \cdot (t/t_{\rm eddy}-4) + 0.04 & \textrm{ for } t/t_{\rm eddy} > 4 \end{array} \right. ,
\label{Bdynamo}
\end{equation}
where $\epsilon_{\rm turb}$ is the turbulent energy density and
$t_{\rm eddy}\equiv 1/|\vec{\nabla} \times \vec{v} | $ is the reciprocal of the vorticity calculated from
the local flow speed.
The fitting function $\phi$ represents the fraction of the
turbulent energy transferred to the magnetic energy via turbulent dynamo, and
it is derived from a magneto-hydrodynamic simulation \citep{ryu2008}.
This model predicts that the magnetic field strength reaches to a few $\microGauss$ in the cluster center
and decreases to $\sim 0.1\ \mu$G toward the cluster outskirts.
This is in a good agreement with the observed magnetic field strength in actual clusters \citep[e.g.,][]
{carilli2002, govoni2004,bonafede11}.}

\subsection{Modeling of CR Electron Spectrum}\label{sec:numerics:spectrum}

For the energy spectrum of CR electrons, 
we first assume that the CR acceleration at shock is described by the test-particle model \citep{drury1983},
since most of the shocks found in clusters are weak (see Introduction).
Then, the momentum distribution of CR protons at the shock position 
is described by the power-law form for $p\ge p_{\rm min}$,
\begin{equation}
f_p(\pnorm) = f_{p0} \times \pnorm^{-q}~,
\end{equation}
where $\pnorm \equiv p/(\mproton c)$ is the dimensionless proton momentum, $\mproton$ is the proton mass,
and $q = 3\sigma / (\sigma-1)$ is the spectral index. 
The shock compression ratio is determined by the sonic Mach number as
$\sigma = {(\gamma +1 )M_s^2}/{[(\gamma -1)M_s^2 + 2]}$.
The normalization factor $f_{p0}$ at each shock zone is determined by setting
the CR energy flux through the shock zone as
\begin{equation}
f_{{\rm CR},p} = v_2 \cdot  \mproton c^2 \int_{\pnorm_\mathrm{inj}} ^{\infty} \left(\sqrt{\pnorm^2+1}-1\right) f_p(\pnorm) d\pnorm^3
\label{fcrp2}
\end{equation}
where $v_2 = {v_1}/{\sigma}$ is the postshock flow speed and $f_{{\rm CR},p}$ is given in
Equation (\ref{fcrp}).
We do {\it not} consider the re-acceleration of pre-existing CRs in this work.

Here, $p_{\rm min}=p_{\rm inj}$ is the injection momentum above which particles can participate 
in the DSA process. 
According to recent PIC and hybrid simulations, both protons and electrons initially can gain energies via SDA
while confined between the shock front and the preshock region by scatterings due to self-generated upstream
waves \citep[][see Introduction]{guo14,park15,caprioli15}.
Particles should have the rigidity ($R = pc/e$) large enough to cross the shock front,
i.e., $p_{\rm inj}\sim 3 p_{\rm th,p} \sim 130 p_{\rm th,e}$, in order to take part in the full
first-order Fermi process.
Here, $p_{\rm th,p}=\sqrt{2m_p k_B T_2}$ is 
{the most probable momentum of thermal protons} of postshock gas with temperature $T_2$, 
while {the most probable momentum of thermal electrons} is $p_{\rm th,e}=(m_e/m_p)^{1/2} p_{\rm th,p}$.
In fact, $p_{\rm inj}$ is expected to depend on the shock Mach number, obliquity angle, and shock
speed.
Here, for the sake of simplicity, we set $\pnorm_\mathrm{inj} = 0.01$ in all shock zones regardless 
of the shock speed and Mach number and the ICM temperature. 

Since the DSA process operates on CR protons and electrons with the same rigidity  
in the same manner, the momentum distribution of primary CR electrons at shocks
should follow that of CR protons, except less efficient injection and radiative cooling \citep[e.g.,][]{kang2011, park15}.
The injection rate of CR electrons is expected be much lower than that of CR protons,
because postshock thermal electrons need to be pre-accelerated from the thermal momentum 
to $p_{\rm inj}$ in order to get injected into the DSA process.
Since the electron pre-acceleration is not yet fully constrained by plasma physics
(despite of recent PIC simulations, see Introduction),
it is often parameterized by the CR electron-to-proton ratio, $K_{e/p}$.
Different types of observations have indicated a wide range of $K_{e/p}\sim 10^{-4} - 10^{-2}$ \citep[e.g.,][]{schlickeiser2002, morlino2009}. 
So we adopt the CR electron momentum distribution of
$f_e(\pnorm) = K_{e/p}\cdot f_p(\pnorm)$ for $p\ge p_{\rm min}$ with $K_{e/p} \sim 0.01$.
{Adopting a higher value of $K_{e/p}$, as suggested in some recent PIC simulations \citep{guo14, caprioli15}, 
will increase the amplitude of $f_e(\pnorm)$ and therefore the synchrotron radiation flux uniformly for all shocks.}

At the same time of gaining energy via the DSA process at shock, CR electrons lose energy via synchrotron emission and IC scattering.
As a results, an equilibrium momentum exists where the momentum gain is balanced by the radiative losses \citep{kang2011}:
\begin{equation}
\pnorm_\mathrm{eq} = {{m_e^2 c v_1}\over {m_p \sqrt{4 e^3 q/27}}} \left[{B_1\over{B_\mathrm{eff,1}^2+B_\mathrm{eff,2}^2}}\right]^{1/2},
\end{equation}
where all the quantities except $\pnorm_\mathrm{eq}$ are expressed in cgs units
and the electron equilibrium momentum is normalized as $\pnorm_\mathrm{eq} =p_\mathrm{eq}/m_pc$.
Here, $B_\mathrm{eff} \equiv (B^2 + B_\mathrm{CBR}^2)^{1/2}$ takes account for the loss due to IC scattering of 
cosmic background photons ($B_\mathrm{CBR} = 3.24 (1+z)^2 \microGauss$)
as well as the synchrotron loss.
Then, the CR electron spectrum {\it at the shock location} has the form of
$f_e(\pnorm) = K_{e/p}\cdot f_p(\pnorm)\cdot \exp \left[ -(\pnorm/\pnorm_\mathrm{eq})^2 \right]$.

As the primary CR electrons are advected downstream in the postshock region,
they also lose energy  
with the cooling time $t_\mathrm{rad} = (0.54 \Gyr) (B_\mathrm{eff,2}/5\microGauss)^{-2} \pnorm^{-1} $.
So the cutoff momentum due to radiative losses decreases with the postshock distance ($d$) away from the shock front as $\pnorm_\mathrm{cut}(d) \propto \pnorm_\mathrm{eq}/d$.
As a result, the downstream, integrated electron spectrum steepens by one power of $p$ as $f_e(\pnorm) \propto \pnorm^{-(q+1)}$ 
for $p\ge \pnorm_\mathrm{br}$,
where the `break' momentum, 
\begin{equation}
\pnorm_\mathrm{br} \approx 0.54 \left( {t_\mathrm{adv} \over 1\Gyr} \right)^{-1} \left( {B_\mathrm{eff,2} \over 5\microGauss} \right)^{-2},
\end{equation}
is the lowest momentum up to which the postshock CR electrons have cooled down.
In our simulation sample, a shock is defined as discontinuity within a grid zone of thickness $\Delta l = 48.8 -$ $ 195.3 \kpch$.
The advection time for CR electrons to pass such thickness is $t_\mathrm{adv} = \Delta l / v_2 \sim 0.1-0.4~\Gyr$ for $v_2\sim 500 \kms$, 
which is of the order of the dynamical time of typical clusters or the merger time.
For the CR electrons emitting at GHz in $\mu$G magnetic field ($\pnorm \sim 7$), the cooling time becomes shorter than
the advection time ($t_\mathrm{rad}(\pnorm) < t_\mathrm{adv}$), so they have cooled down before existing the shock zone.
Hence, we assume that the {\it volume-averaged} spectrum of CR electrons within each shock zone has the following form:
\begin{equation}
f_{e}(\pnorm) = \left\{ \begin{array}{ll}
K_{e/p}\cdot f_{p0} \pnorm^{-q}  & \textrm{ for } \pnorm < \pnorm_\mathrm{br} \\
K_{e/p} \cdot f_{p0} \pnorm_\mathrm{br} \pnorm^{-(q+1)} \cdot \exp \left[ -(\pnorm/\pnorm_\mathrm{eq})^2 \right] & \textrm{ for } \pnorm > \pnorm_\mathrm{br} \\
\end{array} \right. \,.
\label{fep}
\end{equation}

{Note that we made several assumptions and simplifications such as the DSA efficiency in Equation (\ref{fcrp}),
the magnetic field strength, $B$,
in Equation (\ref{Bdynamo}), the electron to proton ratio, $K_{e/p}$, and the volume-averaged electron
spectrum in Equation (\ref{fep}).

Modifications to models for those will lead to rescaling of synchrotron radiation flux.
So these may affect quantitatively some of the shock properties in the flux-limited, mock sample of radio relics,
because a radio relic will consist of multiple shocks in the projected maps (see Section 2.4).
But their impacts on the estimates of derived radio Mach numbers 
would be marginal, so the main results of this work should remain mostly unaffected (see Section 2.5).}

\subsection{Mock Radio/X-ray Maps}

We calculate the radio synchrotron emission at shock zones, using the magnetic field strength given in Equation (\ref{Bdynamo}) and the CR electron spectrum given in Equation (\ref{fep}).
For a single electron with $\pnorm$, the synchrotron power at frequency $\nu$ is given by
$P_{\nu,e} (\pnorm,\theta) = {(\sqrt{3}e^3 B \sin \theta)}/{(m_\mathrm{e} c^2)} \times F\left( {\nu}/{\nu_\mathrm{c}} \right)$,
where $\theta$ is the angle between the electron momentum and magnetic field directions,
$\nu_\mathrm{c} \equiv 3[ (\pnorm (m_p/m_e))^2 + 1] (eB \sin \theta) / (4\pi m_e c)$ is the characteristic frequency,
and $F(x) \equiv x \int_x ^\infty d\xi K_{5/3}(\xi)$ ($K_{5/3}(x)$ is modified Bessel function).
Then, the synchrotron emissivity (in cgs units of erg cm$^{-3}$ s$^{-1}$ Hz$^{-1}$ str$^{-1}$) in shock zone can be estimated 
as the sum of $P_{\nu, e}$ for all CR electrons 
of momentum spectrum $f_e(\pnorm)$ in the zone, assuming random angular distribution (i.e., $\langle \sin \theta \rangle = \pi / 4$),
\begin{equation}\label{eq:synchrotron_emissivity}
j_\nu = \frac{1}{4\pi} \int_{\pnorm_\mathrm{min}}^\infty \frac{\sqrt{3}\pi e^3 B_2}{4 m_e c^2} F\left( \frac{\nu}{\nu_\mathrm{c}(\pnorm)} \right) f_e(\pnorm) d^3 \pnorm \, .
\end{equation}
Note that $j_{\nu}$ is in fact the volume-averaged emissivity, because $f_e(\pnorm)$ in Equation (\ref{fep}) is 
averaged over the volume of the grid zone taking account for radiative cooling in the postshock region. 

The calculation of the bolometric X-ray emissivity due to thermal brem\-sstrahlung in the ICM is
straightforward and can be done with
\begin{equation}
j_{\rm X} \approx (1.2\times 10^{-28}{\rm erg~cm^{-3}~s^{-1}~str^{-1}}) ~T^{1/2} ~ \left({\rho\over m_p}\right)^2, 
\end{equation}
where the He abundance is assumed to be 7.9 \% by number (i.e., the mass fractions of $X=0.76$ and $Y=0.24$ )
and the Gaunt factor $g_{\rm ff} \approx 1.2$ for $T>10^7$K is used.
Note that $j_X$ is calculated at all grid zones, while $j_{\nu}$ only at shock zones.

Mock radio and X-ray maps of each synthetic cluster are constructed by projecting the synchrotron emissivity, $j_\nu$, 
and the bolometric X-ray emissivity, $j_{\rm X}$, along a depth of $D=4 \rtwo$.
The choice of the depth should not affect our results,
since both the synchrotron and X-ray emissions beyond $\rtwo$ from the cluster center are in general negligible.
All line-of-sights (LoSs) are assumed to be parallel to each other,
since the angular sizes of observed clusters and radio relics are usually much smaller than one radian.
Assuming that the system is optically thin,
the synchrotron/X-ray intensities, or surface brightnesses,
are calculated by integrating the synchrotron/X-ray emissivities along each LoS, i.e.,
$I_\nu = \int_\mathrm{LoS}~j_\nu~dl$ and $I_{\rm X} = \int_\mathrm{LoS}~j_{\rm X}~dl$.

Since the spatial distribution of shock zones in clusters is quite complicated, 
they can be translated into radio structures of different morphologies, 
depending on the projection direction \citep[e.g.,][]{vazza2012a, skillman2013}.
So we choose 24 equally-spaced viewing angles for projection,
resulting in 24 realizations of radio/X-ray maps for each of 228 synthetic clusters.
As a result, a total of 5472 radio/X-ray maps are generated and used for the statistics below.

Radio telescopes have finite resolutions,
so in practice they measure the surface brightness convolved with telescope beams.
If a beam has effective angular area $\theta^2$,
then the measured synchrotron flux within the beam is approximately
$S_{\nu_{obs}} \approx I_{\nu} \theta^2 (1+z)^{-3}$, 
where $\nu_\mathrm{obs} = \nu/(1+z)$ is the redshifted observation frequency.
In this study, we adopt $z=0$  
and a beam of $\theta^2 = 15\arcsec \times 15\arcsec$ for synthetic radio observation.
The beam size is chosen to comparable to those of future radio surveys such as those of the Square Kilometre Array (SKA).

Hereafter, we set $\nu_\mathrm{obs}=1.4$ GHz as the representative radio frequency, and present
$j_{1.4}$ and $S_{1.4}$ at the frequency as radio quantities for synthetic observation.
For the calculation of radio spectral index, two frequencies of 142 MHz and 1.4 GHz are used (see the next subsection).

\subsection{``Derived'' Mach Numbers}

With shocks of different Mach numbers and kinetic energy fluxes existing in clusters as mentioned in Introduction,
the radio/X-ray structures projected to the sky may consist of shock surfaces of different characteristics.
So it may not be straightforward to define the properties for observed structures such as radio relics.
In our 3D simulation data cube, each `shock zone' is specified by the sonic Mach number, CR spectrum, and radio emissivity, while each zone is assigned with the gas density, velocity, temperature, magnetic field strength and X-ray emissivity, as described in previous subsections.
In observations (real or synthetic), on the other hand, the physical properties of shocks must be
extracted from 2D projections of radio/X-ray emissivities. 
Here, we describe the quantification of sonic Mach number $M_s$ of shocks associated with 2D projected radio/X-ray structures.

We specify the \emph{derived} Mach numbers in two different ways.
First, the {\it weighted} Mach numbers, $M_{1.4}^\mathrm{w}$ and $M_{X}^\mathrm{w}$, 
are defined as the averages along LoS, weighted by their synchrotron and X-ray emissivities, i.e.,
$M_{1.4}^\mathrm{w} \equiv {\int_\mathrm{LoS} j_{1.4} M_s dl}/{\int_\mathrm{LoS} j_{1.4} dl}$ and
$M_{X}^\mathrm{w} \equiv {\int_\mathrm{LoS} j_{X} M_s dl}/{\int_\mathrm{LoS} j_{X} dl}$, 
respectively.
We consider that these weighted Mach numbers represent the true properties of shocks associated with 2D projected structures.

Second, the {\it observed} Mach numbers, $M_{1.4}^\mathrm{obs}$ and $M_{X}^\mathrm{obs}$, on the other hand,
are designed to mimic the Mach numbers inferred from radio/X-ray observations.
In 2D projected map, a `radio pixel' is the pixel that has at least one shock zone along its LoS and so has $S_{\nu} > 0$.
For each radio pixel, $\Msynctwo$ is calculated from
the integrated spectral index of synchrotron flux, i.e.,
$\alpha \equiv -{d \ln S_\nu}/{d \ln \nu} = (\Msynctwo ^2 + 1)/ ( \Msynctwo ^2 -1)$,
obtained between $\nu_1=142$ MHz and $\nu_2=1.4$ GHz,
following the same way that radio observers usually interpret their observed radio spectra (see Table 1).
Note that the value of $\Msynctwo$ is not sensitive to the choice of low frequency with our model of CR electron spectrum.
However, this practice is justified only when the break momentum of the volume-integrated electron spectrum 
is $\pnorm_\mathrm{br}\ll 1-10$, that is, when the break frequency in the integrated radio spectrum is $\nu_{\rm br}<100$~MHz.

For the calculation of $\Mxtwo$, the X-ray emission-weighted temperature is calculated along each LoS as
$T_X \equiv {\int_\mathrm{LoS} j_{\rm X} T dl}/{\int_\mathrm{LoS} j_{\rm X} dl}$ in mock maps of clusters.
Pixels are tagged as `X-ray shocks', if $| \Delta \log T_X | > 0.11$,
as in the shock identification scheme for 3D volume (see Paper I).
For X-ray shocks, $\Mxtwo$ is estimated from the temperature jump, 
i.e. $T_{X,2}/T_{X,1} = (5 \Mxtwo ^2 -1)(\Mxtwo ^2 + 3)/(16 \Mxtwo ^ 2)$.
The larger of the values of $\Mxtwo$ estimated along the two primary ($x$ and $y$) directions on projected maps is assigned 
as the observed Mach number, i.e.,
$\Mxtwo = \max (M_{X, x}^\mathrm{obs}, M_{X, y}^\mathrm{obs})$.
Again, only X-ray shocks with $\Mxtwo \ge 1.5$ are considered.
Note that a pixel identified as X-ray shock may not be a `radio pixel' and vice versa. 
In fact, only a small fraction ($\sim 7 \%$) of radio pixels are identified as X-ray shocks.
Although the fraction may differ in real observations,
this suggests that X-ray observations could miss a substantial fraction of radio shocks.
Since it is caused by the smoothing in projection,
this problem may be difficult to be overcome, even if the angular resolution of X-ray observation is improved.

\section{Projection Effects on Radio/X-ray Map}\label{sec:projection}

In this section, we examine how the projection affects the radio/X-ray observations of shocks
such as the morphologies and derived shock parameters.
As a representative example, we consider a cluster from $100 \Mpch$ box simulation with $2048^3$ grid zones.
This cluster is identical to the one appeared in the left panels of Figures 1 and also in Figure 7 of Paper I.
It has the X-ray emission-weighted temperature $k_B T_{X, {\rm cl}} \approx 2.7$ keV
and $\rtwo \approx 1.81 \Mpch$.
The shock that produces the largest amount of CR protons in this cluster is an infall shock
with $M_s \approx 5$ and $f_\mathrm{CR} \approx 1.5 \times 10^{47}~\mathrm{erg~s^{-1}} (\Mpch)^{-2}$
(see Paper I).

\fref{map} shows projected maps of the cluster, viewed from four different angles,
zoomed around the area of $r \leq \rtwo$ (dashed circle), where $r$ is the distance from the cluster center
in the projected plane.
Here, the synchrotron flux, $S_{1.4}$, is shown as contours of black solid lines,
which is superposed with the bolometric X-ray surface brightness, $\Ixbol$, in gray-scale.
The lowest contour level for the synchrotron flux is $S_\mathrm{1.4, min} = 10^{-2}\mJy$, 
which could be detected by future radio observatories such as the SKA.
The projection direction of each map is given in terms of polar and azimuthal angles, $\theta$ and $\phi$, respectively.

The four maps in \fref{map} exhibit morphologies of radio structures, distinct from each other.
For instance, the radio map in the upper-left panel seems to show a paired structures on the opposite side of the cluster, which could be interpreted as radio relics due to a pair of merger shocks.
Here, we tag the radio pixels associated with the left structure as ``L-relic'' (red color),
while those associated with the right structure as ``R-relic'' (green color).
In other three panels,
if the shock zone with the largest synchrotron contribution along a given LoS belong to either L-relic or R-relic,
we color the corresponding radio pixel as red or green, respectively. 
In the upper-right and lower-left panels, elongated radio structures with length $ \ga 1-2 \Mpch$
are composed of both red (L-relic) and green (R-relic) pixels.
This illustrates that even a single connected radio structure in the sky may consist 
of a number of disconnected shock surfaces in the real 3D volume.
So it could be misleading, if we try to extract the nature of underlying shocks
only from the morphology or shock parameters inferred from radio observations.
The radio map in the lower-right panel, on the other hand,
has a number of small structures,
and the structure located near the center could be interpreted
as so-called ``radio mini-halo'' \citep[see, e.g.,][for reviews]{feretti2012, bruggen2012}.
Here, we do not intend to address the natures of observed paired structures or halo-like structures
in detail, but we just point that the morphology of observed radio structures critically depends on the
projection of underlying 3D structures.

In \fref{slice}, we compare the spatial distributions of $T$, $\rho$, and $M_s$ in a slice (left panels)
with those of $T_X$, $\Msynctwo$, and $\Mxtwo$ in 2D projection (right panels).
The slice passes through $0.37 \Mpch \sim 0.2 \rtwo$ away from the cluster center,
and contains the most radio-luminous shock zones for the L and R-relics,
which are projected at $(-0.4 \Mpch,$ $1.0 \Mpch)$ and $(0.4 \Mpch,$ $0.1 \Mpch)$ in the 2D maps, respectively. 
They are identified as infall shocks with $M_s \approx 5.7$ for L-relic and $M_s \approx 3.9$ for R-relic.
The right panels are from the same map as that for the upper-left panel of \fref{map},
and the contours are drawn with the level of $S_\mathrm{1.4, min}$.
For $\Msynctwo$, only pixels with $S_{1.4}\ge S_\mathrm{1.4, min}$ are plotted,
while for $T_X$ and $\Mxtwo$, only pixels with  $I_{\rm X} \ge I_{\rm X, min}=10^{-10} {\rm erg~s^{-1}~cm^{-2}~str^{-1}}$ are plotted.
Note that the bolometric X-ray surface brightness in outskirts near $r\sim \rtwo$ ranges typically
$1-10\times 10^{-9} {\rm erg~s^{-1}~cm^{-2}~str^{-1}}$ in observed clusters \citep[e.g.,][]{ettori09}.

Comparison of $M_s$ in the slice map (left-bottom) and $\Msynctwo$ in the 2D map (right-middle) indicates that
the Mach number and location of shocks with $M_s \ga 3$ agree well in the two distributions,
especially for the shocks associated with L-relic.
This is because the radio spectrum in projected maps is governed by one or a few radio-luminous shock zones
with high Mach numbers in a given LoS.
So $\Msynctwo$ derived from radio observations could be a good proxy for the Mach number of radio-luminous shocks,
implying that the properties of such shocks could be extracted reasonably well from radio observations.

On the other hand, there are noticeable differences between the distributions of $T$/$M_s$ in the slice maps
(left-top/bottom) and those of $T_X$/$\Mxtwo$ in the 2D maps (right-top/bottom).
As seen in top panels, the distribution of $T_X$ is smoother than that of $T$,
partly because the contribution from the X-ray bright ICM dominates over the emissions 
from the WHIM in surrounding filaments, and also because the projection (integration along LoSs) inevitably irons out any sharp features.
The temperature jump across a shock in the $T_X$ map looks reduced. 
We find that $\Mxtwo$ derived from X-ray observations tends to underestimate the
actual shock Mach number (see the next chapter).

We further examine in detail the region near L-relic, where the slice maps indicate a shock of
$M_s \approx 5 - 6$ formed by the infall of the WHIM along a filament.
There is $\sim 100 \kpch$ of positional shift between the locations of shocks in the $\Msynctwo$ (right-middle) and $\Mxtwo$ (right-bottom) maps.
This is because $\Msynctwo$ picks up the infall shock, while due to the dominant contribution of X-ray from the hot and dense ICM along the LoS, $\Mxtwo$ picks up a foreground shock with smaller Mach number formed in the ICM (sse below).
This may explain spatial offsets between the shock surfaces inferred from radio and X-ray observations
in some radio relics 
\citep[e.g.,][]{akamatsu2013,ogrean2013b}.

\fref{3d} shows the distributions of radio and X-ray emissivities, $j_{1.4}$ and $j_X$,
as functions of $M_s$ for shocks within $r \leq \rtwo$ in our simulated clusters.
Both distributions have rather large spreads, but the following points are clear,
Firstly, $j_X$ decreases with increasing $M_s$, 
while $j_{1.4}$ increases with $M_s$, peaks at $M_s \sim 5$ and 
decreases slightly for higher $M_s$.
These behaviors can be understood as follows.
Weaker shocks which tend to form in hot and dense regions near the cluster core produce more $j_X$,
while shocks formed in cluster outskirts with $M_s \sim$ several accelerate CRs and produce $j_{1.4}$ most efficiently.
Secondly, on average, $j_X$ varies over two orders of magnitude, while $j_{1.4}$ varies over 10 orders of magnitude.
Although both radio/X-ray surface brightnesses on projected maps are dominated
by contributions from a small number of bright shock zones along each LoS, 
the tendency is much stronger in the case of radio.

\section{Shock Properties Derived from Radio/X-ray Maps}\label{sec:stats}

\subsection{Shock Pixels in Projected Maps}\label{sec:stats:point}

In this section, we first examine the surface brightnesses and derived Mach numbers of pixels with shocks, i.e
$\Ssync$, $M_{1.4}$, $I_X$ and $M_X$, obtained in projected radio/X-ray maps.
As noted in Section 2.5, radio pixels which contain shocks in radio maps may not be identified as shocks
in X-ray maps, and vice versa. 
So when we study a correlation between any two quantities, for instance, $\Msynctwo$ versus $\Mxtwo$,
we use a subset of pixels in which both quantities are specified.

\fref{xlum_sync_mach_all} displays the relations between $\Ssync$ vs $I_X$, $\Msynctwo$ vs $\Ssync$,
and $\Mxtwo$ vs $I_X$ for pixels within $r \leq \rtwo$ from the cluster center.
The bottom panel shows that bright X-ray shocks with large $I_X$ tend to be weak with small $\Mxtwo$;
the brightest X-ray shocks have $\Mxtwo \la 2$.
The middle panel, on the other hand, shows that the synchrotron flux $\Ssync$ of radio pixels increases with $\Msynctwo$ up to $\sim 4$ and then decreases for larger $\Msynctwo$.
These behaviors should be guessed from those of $j_{1.4}$ and $j_X$ shown in \fref{3d}.
As a result, the correlation between $\Ssync$ and $\Ixbol$ turns out to be rather poor with wide variations
in the top panel.
What is clear is that the shocks which are brightest in radio are not necessarily brightest in X-ray.
As a matter of fact, brightest X-ray shocks have modest radio brightnesses, and vice verse.

\fref{machs_all} displays the relations among mock-observed Mach numbers, $\Msynctwo$ and $\Mxtwo$,
and weighted Mach numbers, $\Msyncthree$ and $\Mxthree$, for pixels within $r \leq \rtwo$.
As noted in the previous chapter, the X-ray Mach numbers tend to be smaller than the radio Mach numbers (upper panels).
Again, this is mainly because X-ray observations incline to pick up weaker shocks than radio observations 
along given LoSs.
The correlation between $M_{1.4}$ and $M_X$ is rather poor;
the Pearson's linear correlation coefficient of $\Msynctwo$ and $\Mxtwo$ is $r(\log \Msynctwo, \log \Mxtwo)=0.11$ (upper-left panel) and that of $\Msyncthree$ and $\Mxthree$ is $r(\log \Msyncthree, \log \Mxthree) = 0.22$
(upper-right panel).
So it could be misleading if $\Mxtwo$ is guessed from $\Msynctwo$, and vice versa.
It is expected that $M_{\rm obs}$ and $M_{\rm w}$ for the same band are correlated better.
We find that $\Msynctwo$ is often smaller than $\Msyncthree$ as shown in the lower-left panel,
but the correlation between them is quite good with $r(\log \Msynctwo, \log \Msyncthree) = 0.77$.
The correlation between $\Mxtwo$ and $\Mxthree$ in the lower-right panel, on the other hand, is still poor
with $r(\log \Mxtwo, \log \Mxthree) = 0.24$.
This demonstrates that the smoothing in projection would make the estimation of shock properties harder in X-ray observations than in radio observations.

In \fref{rad_all}, we examine the radial distributions of observable properties of shock pixels.
Both $\Msynctwo$ and $\Mxtwo$ tend to increase with radius,
and $\Msynctwo$ shows a larger variance than $\Mxtwo$.
Shocks with X-ray Mach number $\Mxtwo \ga$ a few would be rare within $r \leq \rtwo$,
while shocks with radio Mach number $\Msynctwo$ up to several could be found.
Radio bright shocks, for instance, those with $\Ssync\ge S_\mathrm{1.4, min} = 10^{-2}\mJy$,
are found mostly in $r \ga 0.2 \rtwo$.
This suggests that future radio observations could detect many more shock structures in cluster outskirts.
On the other hand, X-ray bright shocks are preferentially located and so found close the center.
The bolometric X-ray surface brightness of X-ray shocks follows $\Ixbol \propto (r/\rtwo)^{-\beta}$ with $\beta\approx 3-5$
for $r/\rtwo > 0.3$, which is consistent with the observed X-ray profiles in cluster outskirts \citep[e.g.,][]{ettori09}.

\subsection{Radio Relics in Projected Maps}

We build up a catalog of ``synthetic radio relics'' by
finding connected structures in our radio maps, which meet the following conditions.
(1) All pixels within structures have $\Ssync \geq S_\mathrm{1.4, min}$.
(2) A structure has at least five pixels.
Note that the area of a single pixel is $\Delta A=(L/N_g)^2$ ($N_g=1024$ or $2048$).
So five pixels corresponds to, for instance, $A_\mathrm{min} \approx 0.05 (\Mpch)^2$ in simulations of $L = 100 \Mpch$ box with $1024^3$ grid zones.
Using these conditions, we have 9583 radio relics samples:
4626 from simulations of $L = 100 \Mpch$ with $1024^3$ grid zones,
4850 from simulations of $L = 200 \Mpch$ with $1024^3$ grid zones,
and 107 from simulation of $L = 100 \Mpch$ with $2048^3$ grid zones.

We then assign the following quantities to synthetic radio relics.
The average distance from the cluster center, $\rrelic$, is defined by
the synchrotron-weighted average of the distance of the pixels that belong to a radio relic,
$\rrelic \equiv {\sum_{\rm pixel} r \Ssync}/{\sum_{\rm pixel} \Ssync }$,
where $r$ is the distance of pixels from the cluster center.
The derived Mach numbers of radio relics are defined by the synchrotron/X-ray weighted averages 
of the corresponding derived Mach numbers of pixels, $\Msynctwo$, $\Msyncthree$, $\Mxtwo$, and $\Mxthree$.
For example, $M_{1.4,\rm relic}^{\rm obs} \equiv {\sum_{\rm pixel} \Msynctwo \Ssync}/{\sum_{\rm pixel} \Ssync }$
and  $M_{X,\rm relic}^{\rm obs} \equiv {\sum_{\rm pixel} \Mxtwo I_X}/{\sum_{\rm pixel} I_X }$.
The number of pixels with $\Mxtwo$ could be smaller than that with $\Msynctwo$,
since $\Mxtwo$ is specified only at the pixels identified as X-ray shocks.
In fact, about 40\% of sample radio relics contain no pixel with $\Mxtwo$, so the corresponding X-ray Mach number is not assigned to them.
The synchrotron power at 1.4 GHz and bolometric X-ray luminosity of a radio relic are calculated with
$\Psync = 4\pi \Delta A \sum_{\rm pixel} \Ssync$ and $\Lxbol = 4\pi \Delta A \sum_{\rm pixel} I_X $, respectively.

\fref{machs_relic} displays the relations among the derived Mach numbers for sample radio relics.
The relations follow and so look similar to those for pixels in projected maps shown in \fref{machs_all},
but the correlations are tighter with
$r(\log \Msynctworelic,$ $\log \Mxtworelic)$ $= 0.65$,
$r(\log \Msyncthreerelic,$ $\log \Mxthreerelic)$ $= 0.64$,
$r(\log \Msynctworelic,$ $\log \Msyncthreerelic)$ $= 0.99$, and
$r(\log \Mxtworelic,$ $\log \Mxthreerelic$)$ = 0.38$.
Especially, the correlation between $\Msynctworelic$ and $\Msyncthreerelic$ is very good,
indicating that the Mach number estimated from radio spectral index
would be a fair representation of suitably averaged Mach number of shocks associated with radio relics.
Most of our sample radio relics have $\Msynctworelic \ga 2.5$.
This is partly because with the model CR acceleration efficiency we employ \citep{kang2013},
the amount of CR electrons emitting synchrotron is very small for weak shocks with $M_s \la 2.5$ (see Figure 3 of Paper I).
If the CR electron acceleration at weak shocks would be more efficient than in our model,
radio relics with smaller $\Msynctworelic$ could be more common.
As noted in Introduction, Merger shocks are expected to have mostly $M_s \la 3$.
In fact, we find that our synthetic radio relics normally involve projections of multiple shocks along LoSs, resulting in different morphologies for different viewing angles (see \fref{map}).
In $\sim 40\%$ of sample radio relics, the brightest pixels with largest $\Ssync$ include infall shocks along LoSs,
suggesting that infall shocks may account for some radio relics with flat spectra.
This seems reasonable in the sense that infall shocks could be the major sources of CRs in clusters, as discussed in Paper I.

\fref{rad_relic} displays the radial distributions of $\Msynctworelic$, $\Mxtworelic$, 
$\Psync$, and $\Lxbol$.
Upper panels show that most of our synthetic radio relics are found at $\rrelic \ga 0.2 \rtwo$.
Although there are wide variations, $\Msynctworelic$ and $\Mxtworelic$ tend to increase toward outskirts,
and on average the radio Mach number is larger than the X-ray Mach number.
The lower-left panel shows that the radio power at a given radial distance can vary widely,
yet most powerful radio relics are found at $0.3\la \rrelic/\rtwo \la 0.7$.
This results mainly from the combined effects of
the radial distribution of shock kinetic energy flux and $M_s$,
the strong dependence of CR acceleration efficiency on $M_s$,
and the geometrical increase of relic surface area toward outskirts \citep[see also][]{vazza2012a}.
The lower-right panel shows a strong radial decline of the bolometric X-ray luminosity of 
sample radio relics (not including the contributions from the background ICM) toward outskirts.
This reflects mostly the radial decreases of gas density and temperature;
the effect of the increase of the Mach number and relic surface area is not large here.

\subsection{Comparison with Observational Data}

We compare our synthetic radio relics with observed radio relics.
Table 1 lists the properties of the observed radio relics we use;
here, $\zcl$ is the redshift of associated clusters, $\lrelic$ is the largest linear scale on the sky,
and $\alpha$ is the integrated radio spectral index. 
They are chosen from \citet{vanweeren2009} and \citet{bonafede2012} with the following criteria;
(1) $\zcl \la 0.2$ (our radio relic sample is constructed at $z=0$),
(2) $\lrelic \ge A_\mathrm{min}^{1/2}$ with $A_\mathrm{min} = 0.05 (\Mpch)^2$,
and (3) both $\Psync$ and $\alpha$ available.
Note that $\alpha$ in Table 1 was calculated between $74$ MHz and $1.4$ GHz.
Although some observations suggested that the radio spectra of radio relics could have curvatures and
$\alpha$ may increase at high frequencies $\ga 10$ GHz
\citep[e.g.,][]{vanweeren2009,stroe13, trasatti15}, here we ignore the effect.

\fref{relic} shows $\Psync$ vs $\rrelic$ and $\alpha$ for synthetic (colors) and observed (filled squares) radio relics.
The distribution of $\Psync$ of observed radio relics seems to fall within the range predicted by our synthetic observation.
On the other hand, there are noticeable differences in the distributions of $\rrelic$ and $\alpha$ between synthetic and observed radio relics.
First, four observed radio relics are found with $\rrelic \leq 0.43$ Mpc among twelve, while
the fraction of synthetic radio relics with $\rrelic \leq 0.43$ Mpc is very small, less than $\sim 1 \%$.
Second, while the integrated spectral index of synthetic radio relics lies $1.0 < \alpha \la 1.38$
(corresponding to $ M_s \ga 2.5$), some of observed radio relics have steeper spectral indices.
For example, the radio relics found at Abell 548b B and Abell S753 have $\alpha \ga 2$ ($M_s \la 1.7$),
and the radio relic at Abell 2345W has $\alpha\approx 1.5$ ($M_s \approx 2.1$). 
The discrepancies could be partly because weak shocks close to cluster core might be
under-represented in our structure formation simulations due to the limited spatial resolution and omission of
some non-gravitational processes.
But it could be also possible that weak shocks accelerate CR electrons more efficiently than we assume here,
as hinted in recent PIC simulations \citep[see, e.g.][]{guo14,park15}.

{On the other hand,
the re-acceleration of ``pre-existing'' CR electrons, which is not considered in this study,
could play an important role in the production of CR electrons at weak shocks with $M_s \la 2$
\citep[e.g.][]{kang2012, pinzke2013}.
Especially, if fossil relativistic electrons are present in the form of isolated clouds, instead of being
spread throughout the general ICM space, the relative rareness of radio relics can be understood, 
as discussed in the Introduction \citep{shimwell15,kangryu2015}.
In such a scenario, `lighting-up' of radio relics is governed mainly by the occasional encounters of
ICM shocks with fossil electron clouds, instead of the Mach-number-dependent DSA
efficiency and the shock kinetic energy flux. As a result, the frequency and radio luminosity of radio
relics may depend less sensitively on the shock Mach number, and much weaker shocks 
could be turned on as radio relics with steeper spectral index.
These features are more consistent with the observational data in Figure 9.
Moreover, only a small fraction of ICM shocks
become radio sources for the period much shorter than the cluster dynamical time sale \citep{kangryu2015}.
So the model naturally explains why radio relics are rare, while structure formation simulations predict 
shocks should be ubiquitous in the ICM \citep[e.g.][]{ryu2003,vazza2012a}.
The exploration of this scenario requires extensions of the models we employ here,
so beyond the scope of this paper.} 

\section{Summary}\label{sec:summary}

The existence of shocks in clusters of galaxies has been established through
observations of radio relics due to synchrotron emission from shock-accelerated CR electrons
\citep[e.g.,][]{bonafede2012} as well as through X-ray observations of shock discontinuities
\citep[e.g.,][]{markevitch2007}.
However, it is not clear whether the properties of shocks inferred from radio and X-ray observations represent
their true nature, because we can only measure the quantities projected onto the sky, integrated along LoSs.
{In fact, for a few relics such as the Tooth brush relic and the radio relic in A2256,
radio and X-ray observations have reported inconsistencies in quantities such as shock Mach number and position \citep{akamatsu2013, ogrean2013b,trasatti15}.}

{In this paper, we explored a scenario in which electrons freshly injected and accelerated 
to high energies via DSA process at the structure formation shocks produce radio relics.
We constructed synthetic maps of galaxy clusters in radio and X-ray by
employing the cosmological hydrodynamic simulation data reported earlier in Paper I.}
First, the volume-averaged synchrotron emissivity at shock zones was calculated by
adopting the CR electron acceleration efficiency based on a DSA model \citep{kang2013} 
and magnetic fields based on a turbulent dynamo model \citep{ryu2008}.
The bolometric X-ray emissivity at grid zones was also calculated using the gas density and temperature
from simulations.
Then, mock maps of the synchrotron and X-ray surface brightnesses in 2D projection were produced by integrating volume emissivities along LoSs for simulated clusters.
In the synthetic maps, the properties of shocks and radio relics, 
such as the shock Mach number and location, were examined in detail.

The main findings can be summarized as follows.

{1) In most cases, radio and X-ray shocks in 2D maps are the outcomes of projection of multiple shock 
surfaces along LoSs, 
because shocks are abundant in the ICM with a mean separation of shock surfaces, $\sim 1\Mpch$.
As a result,
the morphology of shock distributions in 2D maps depends on the projection direction; for the same cluster, very different morphologies may turn out for different viewing angles.}

{2) Synchrotron emissivity depends sensitively on the shock Mach number,
especially with our DSA model for the CR electron acceleration,
while bremsstrahlung emissivity depends on the gas density and temperature.
Hence, radio observations tend to pick up shocks of $M_s \sim$ a few to several along a LoS,
while X-ray observations preferentially select weaker shocks ($M_s \la 2$) with high density and temperature.
Consequently, the properties of shocks in 2D projected maps could be different in radio and X-ray observations.
The shock Mach number estimated with X-ray observations tends to be smaller than that from radio observations,
if a radio relic consists of multiple shocks along LoSs.
In addition, the location of shocks from X-ray observations could be shifted with respect to that 
from radio observations.}

{3) For radio relics, the Mach number estimated from the radio spectral index seems to be a fair
representation of suitably averaged Mach number of shock surfaces associated with them.
On the other hand, the discontinuities in X-ray temperature tend to be smeared due to projection effects
including possible multiple shocks and multiple ICM components, 
resulting in the X-ray Mach number lower than the real Mach number of the associated shock.}

{4) When the properties of our synthetic radio relics are compared with those of observed radio relics,
there are clear differences in the statistics of radial location and spectral index;
more radio relics have been observed closer to the cluster core and with steeper spectral indices
than our synthetic observation predicts.
This discrepancy may imply that weak shocks in high beta plasmas in fact accelerate CR electrons 
more efficiently than we model here \citep[see, e.g.][]{guo14,park15}.
Alternatively, as shown in the previous studies of \citet{kang2012} and \citet{pinzke2013},
the re-acceleration of pre-existing electrons in the ICM may enhance the production 
of CR electrons at weak shocks with $M_s \la 2$.
Finally, we could also conjecture that radio relics might be activated only when shocks
encounter clouds, which contain fossil electrons either accelerated earlier by shocks/turbulence or
left over from old radio jets
\citep[see, e.g.][]{shimwell15, kangryu2015}. 
This model may explain how very weak shocks can turn on the synchrotron emission and
why radio relics are rare relative to putative shocks in galaxy clusters.}



\begin{acknowledgements}

{The authors thank the anonymous referee for his/her thorough review and constructive
suggestions that lead to an improvement of the paper.}
SEH was supported by the National Research Foundation of Korea through grant 2007-0093860.
HK was supported by Basic Science Research Program through the National Research 
Foundation of Korea(NRF) funded by the Ministry of Education (2014R1A1A2057940).
The work of DR was supported by the year of 2014 Research Fund of UNIST (1.140035.01).

\end{acknowledgements}

\clearpage

\begin{deluxetable} {lcccccl}
\tablecaption{Properties of Observed Radio Relics}
\tablehead{
\colhead {Source Name} & \colhead{$\zcl$} & \colhead{$\Psync$ } & \colhead{$\rrelic$} & \colhead{$\lrelic$} 
& \colhead{ $\alpha$ } & \colhead{Reference} \\
\colhead { } & \colhead { } & \colhead { $[10^{24}~\mathrm{W~Hz^{-1}}]$ } & \colhead { [kpc] } 
& \colhead {[kpc]} & \colhead { } & \colhead{}
}
\startdata
Abell 2256 & 0.058 & 3.95 & 300 & 1100 & 1.2 & \cite{clarke2006}\\
Abell 1240 N & 0.159 & 0.427 & 700 & 650 & 1.2 & \cite{bonafede2009}\\
Abell 1240 S & 0.159 &  0.730 & 1100 & 1250 & 1.3 & \cite{bonafede2009}\\
Abell 2345 E & 0.177 & 2.62 & 890 & 1500 & 1.3 & \cite{bonafede2009}\\
Abell 2345 W & 0.177 & 2.83 & 1000 & 1150 & 1.5 & \cite{bonafede2009}\\
Abell 3667 & 0.056 & 17.4 & 1950 & 1920 & 1.1 & \cite{roettgering1997}\\
Abell 548b B & 0.042 & 0.250 & 430 & 370 & $> 2.0$ & \cite{feretti2006}\\
Coma 1253+275 & 0.023 & 0.284 & 1940 & 850 & 1.18 & \cite{giovannini1991}\\
Abell 2163 & 0.203 & 2.23 & 1550 & 450 & 1.02 & \cite{feretti2004}\\
Abell S753 & 0.014 & 0.205 & 410 & 350 & 2.0 & \cite{subrahmanyan2003}\\
Abell 115 & 0.197 & 16.7 & 1510 & 1960 & 1.1 & \cite{govoni2001}\\
Abell 610 & 0.095 & 0.444 & 310 & 330 & 1.4 & \cite{giovannini2000}\\ 
\enddata
\end{deluxetable}

\clearpage

\begin{figure}
\plotone{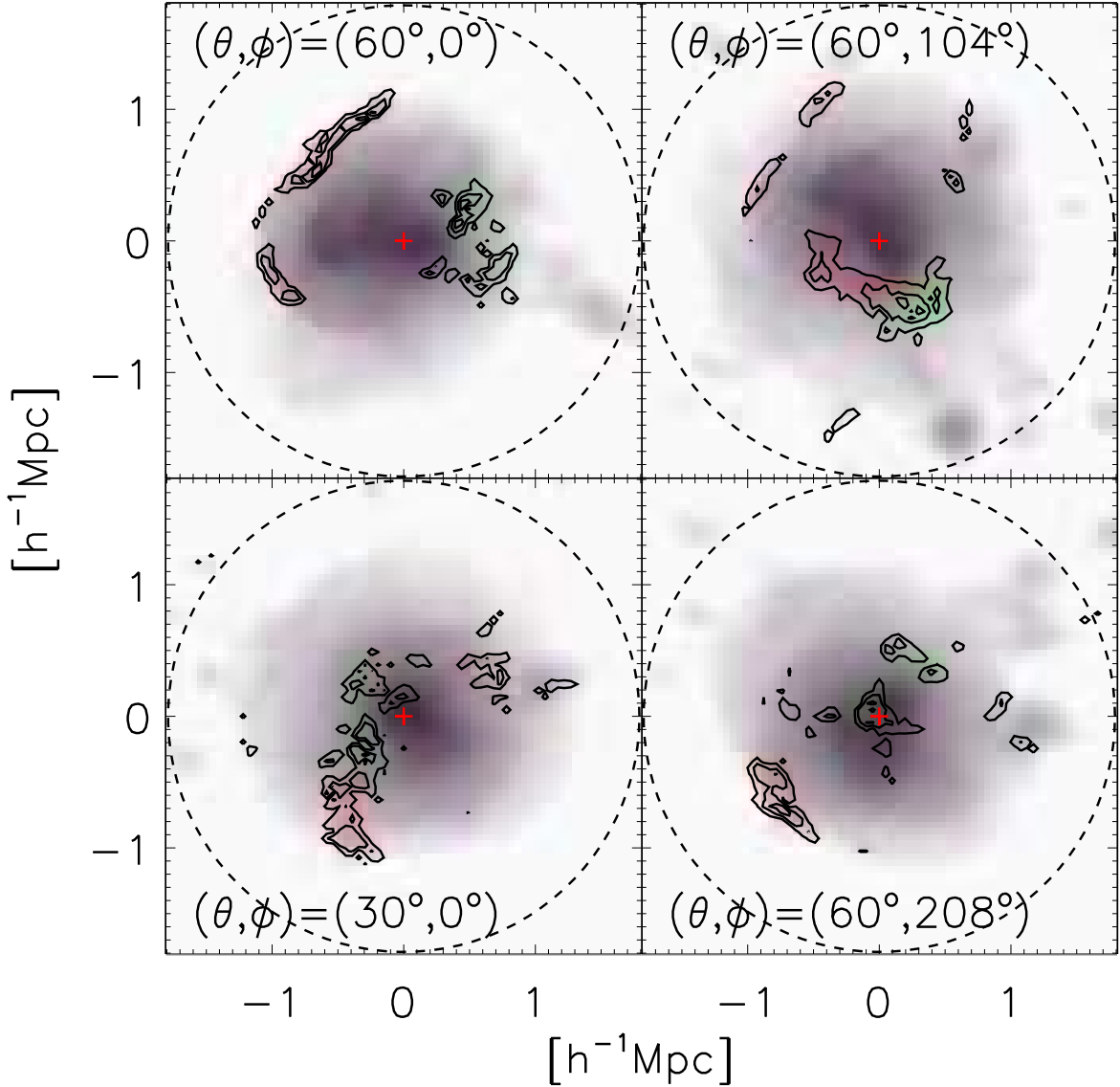}
\caption{Projected maps of a simulated cluster with $k_B T_{X, {\rm cl}} \approx 2.7$ keV at $z=0$,
viewed from four different directions with $\theta$ and $\phi$ listed.
Contours show the synchrotron flux at 1.4 GHz, $S_\mathrm{1.4}$, in logarithmic scale, 
from $10^{-2}$ to $1 \mJy$ with interval of $10^{0.5}$, where a beam of $\theta^2 = 15\arcsec \times 15\arcsec$ is adopted.
Grey-scale maps show the bolometric X-ray surface brightness, $I_X$, in logarithmic scale, 
from $10^{-8}$ (white) to $10^{-2}~\mathrm{ergs~ s^{-1}~cm^{-2}~str^{-1}}$ (black).
Dashed circles show $\rtwo$.
Red and green colors differentiate the shock membership associated with the left and right ``radio relics'', respectively, in the upper-left panel.
The red crosses mark the X-ray center of the cluster.}\label{fig:map}
\end{figure}

\clearpage

\begin{figure}
\vspace{-0.6cm}
\begin{center}
\includegraphics[scale=0.5]{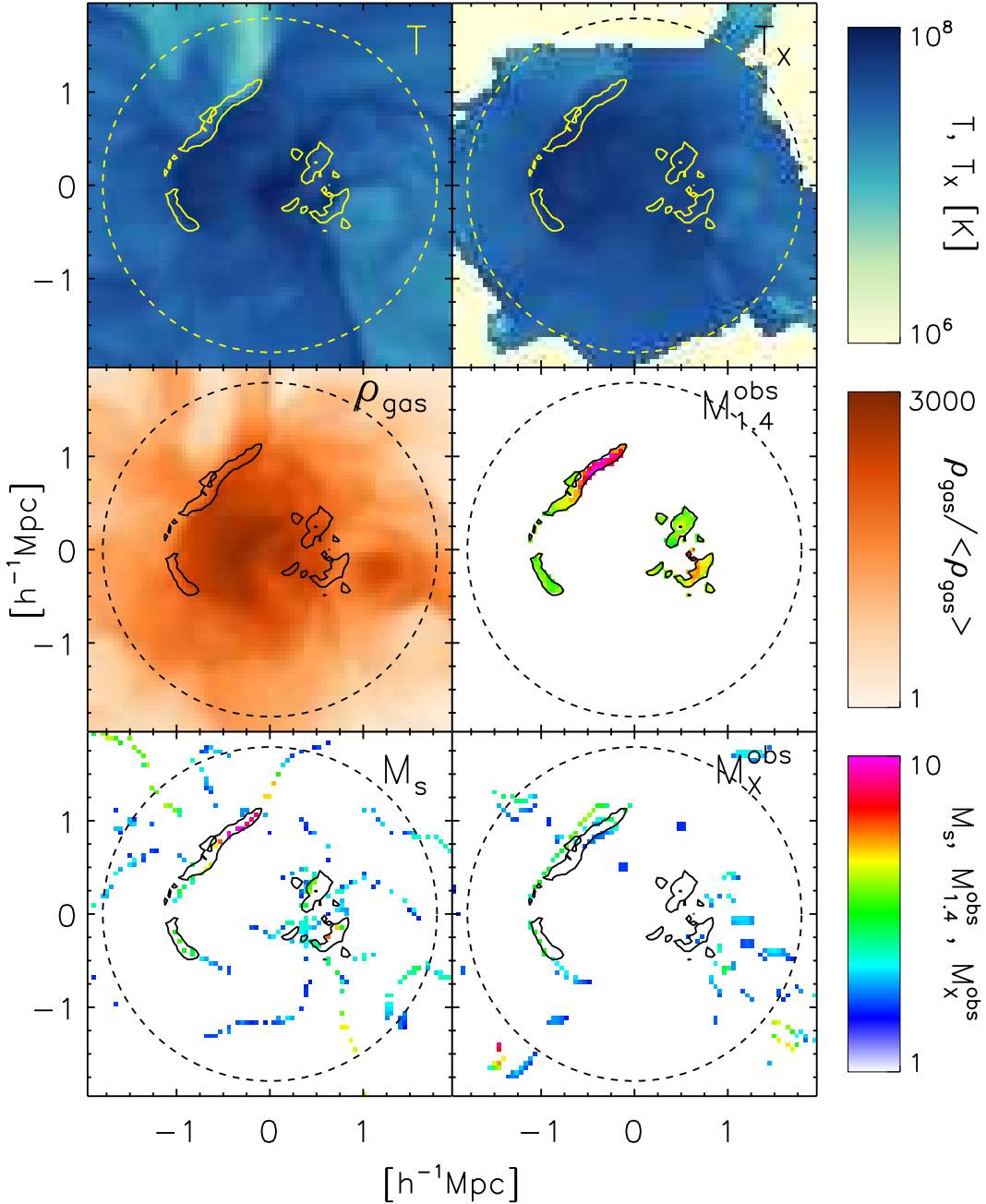}
\end{center}
\vspace{-0.8cm}
\caption{Left panels: Distributions of gas temperature $T$ (top), gas density $\rho$ (middle), 
and shock Mach number $M_s$ (bottom) in a slice which passes through $\sim 0.2 \rtwo$ away from the center
of the same cluster as in \fref{map}.
Right panels: Projected maps of X-ray emission-weighted temperature $T_X$ (top), 
Mach number estimated from the radio spectral index, $\Msynctwo$ (middle),
and Mach number of X-ray shocks detected in the $T_X$ map, $\Mxtwo$ (bottom),
with the same viewing angle as the upper-left panel of \fref{map}.
The yellow (top panels) and black (middle and bottom panels) contours correspond to $S_\mathrm{1.4,min} = 10^{-2} \mJy$.
For $T_X$ and $\Mxtwo$, the pixels with $I_{\rm X}\ge I_{\rm X,min}= 10^{-10}~\mathrm{ergs~ s^{-1}~cm^{-2}~str^{-1}}$ are shown,
while for $\Msynctwo$, the pixels with $S_{1.4}\ge S_\mathrm{1.4,min}$ are shown.
Dashed circles show $\rtwo$.}\label{fig:slice}
\end{figure}

\clearpage

\begin{figure}
\plotone{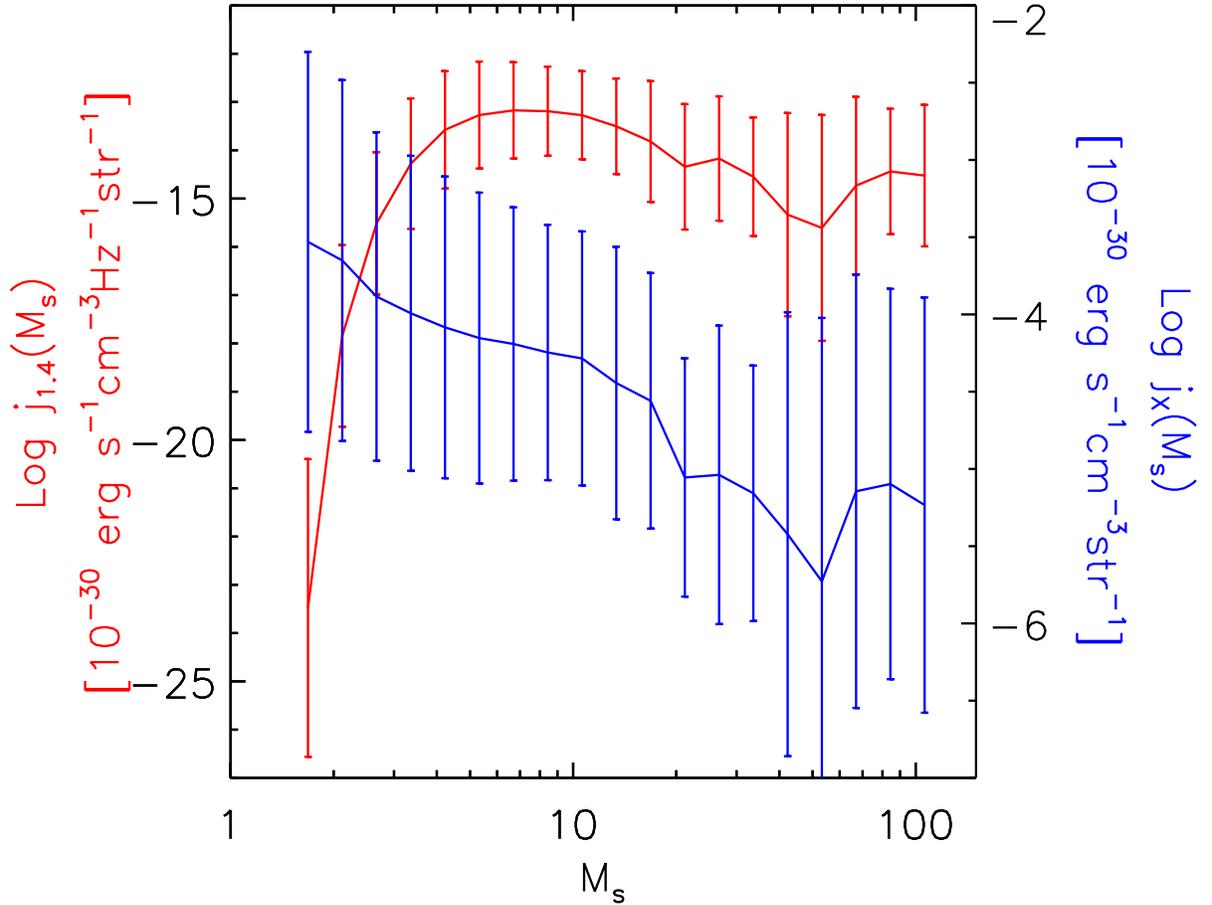}
\caption{Synchrotron emissivity at 1.4GHz, $j_{1.4}$, (red) and bolometric X-ray emissivity, $j_X$, (blue)
vs Mach number $M_s$ for shock zones within $r \leq \rtwo$ from the center
in simulated clusters.
Connected lines and error bars are averages and standard deviations of $\log j_{1.4}$ and $\log j_X$
within Mach number bins of $[\log M_s, \log M_s + \Delta \log M_s]$.}\label{fig:3d}
\end{figure}

\clearpage

\begin{figure}
\vspace{-0.8cm}
\begin{center}
\includegraphics[scale=0.48]{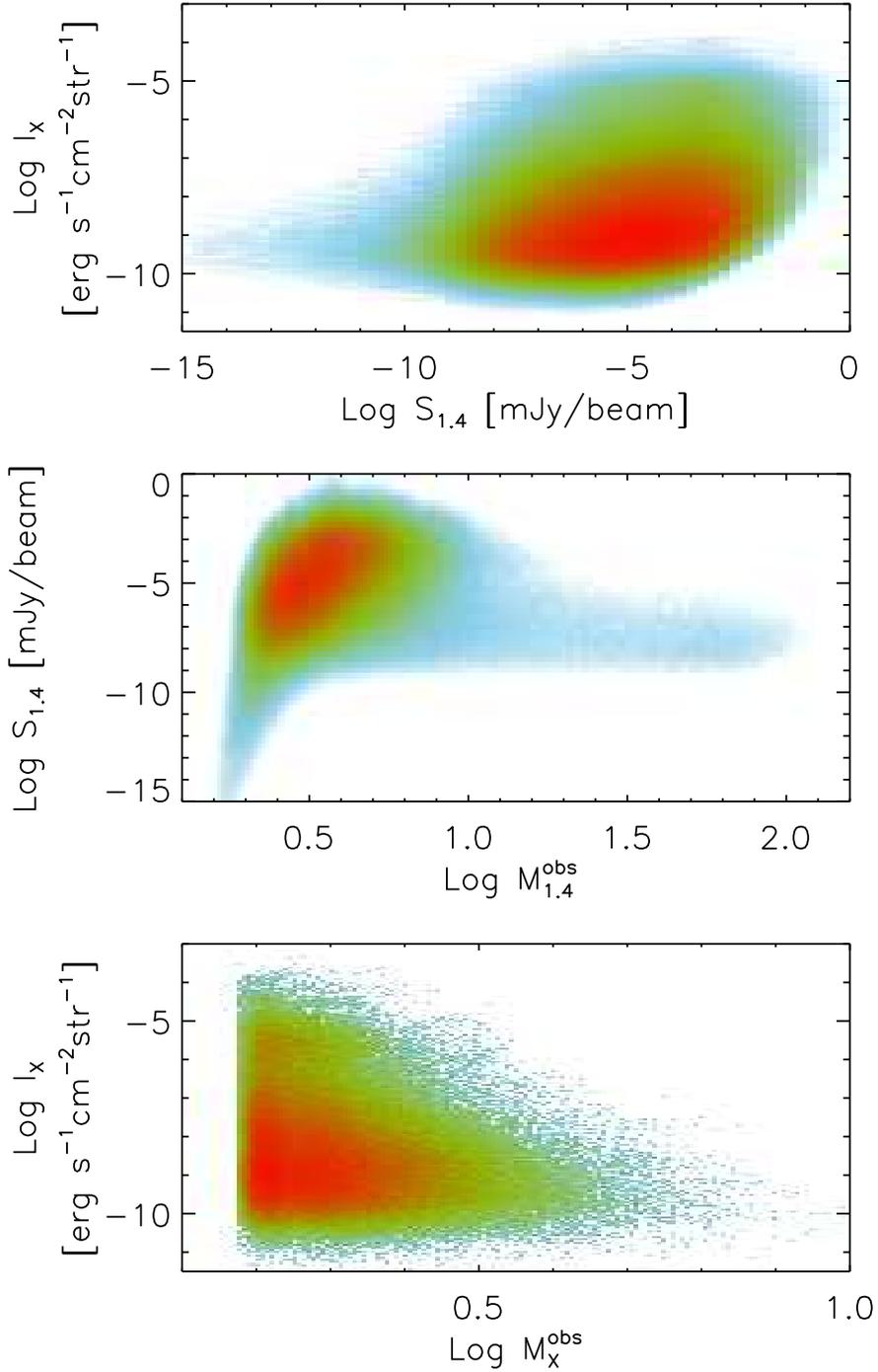}
\end{center}
\vspace{-0.9cm}
\caption{$\Ssync$ vs $I_X$, $\Msynctwo$ vs $\Ssync$, and $\Mxtwo$ vs $I_X$ for shock pixels within
$r \leq \rtwo$ from the cluster center in projected maps.
Here, $\Ssync$ is the synchrotron flux at 1.4 GHz, $I_X$ is the bolometric X-ray surface brightness,
$\Msynctwo$ is the Mach number estimated from the radio spectral index, 
and $\Mxtwo$ is the Mach number of X-ray shocks detected in the $T_X$ map.
Colors code the relative frequency, ranging from $10^{-2.5}$ (blue) to 1 (red) in the top and middle panels
and from $10^{-3.5}$ (blue) to 1 (red) in the bottom panel.}\label{fig:xlum_sync_mach_all}
\end{figure}

\clearpage

\begin{figure}
\plotone{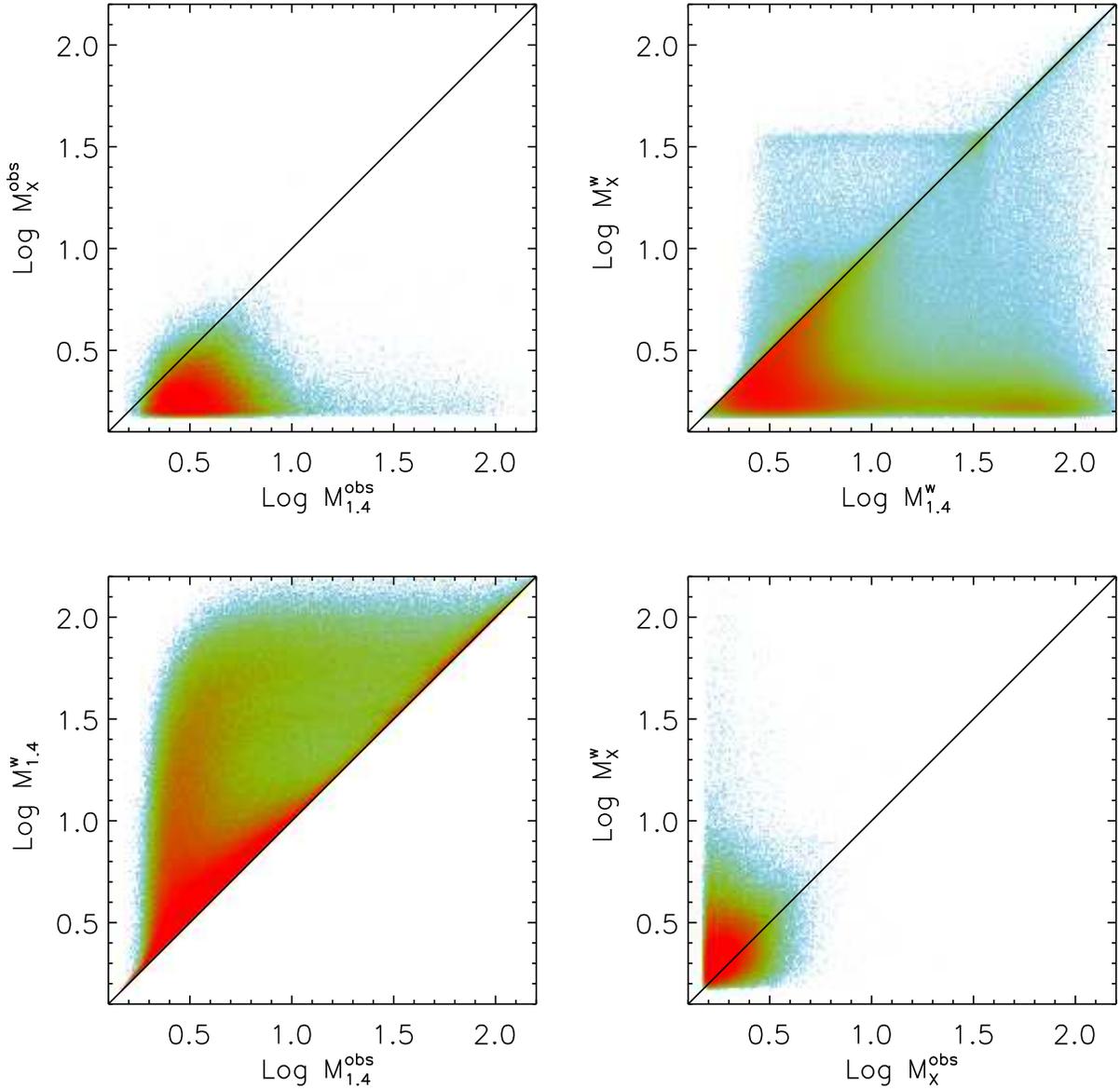}
\caption{Relations among observed Mach numbers, $\Msynctwo$ and $\Mxtwo$, and weighted Mach numbers, $\Msyncthree$ and $\Mxthree$, (see the text for definitions) for shock pixels within $r \leq \rtwo$ from
the cluster center in projected maps.
Colors code the relative frequency, ranging from $10^{-2.5}$ (blue) to 1 (red) in the upper-left and lower-right panels, from $10^{-4}$ (blue) to 1 (red) in the upper-right panel, and from $10^{-3}$ (blue) to $1$ (red) in the lower-left panel.
Pearson's linear correlation coefficients between two derived Mach numbers at each panel are 
0.11 (upper-left), 0.22 (upper-right), 0.77 (lower-left), and 0.24 (lower-right).
Black lines show the perfection correlation.}\label{fig:machs_all}
\end{figure}

\clearpage

\begin{figure}
\plotone{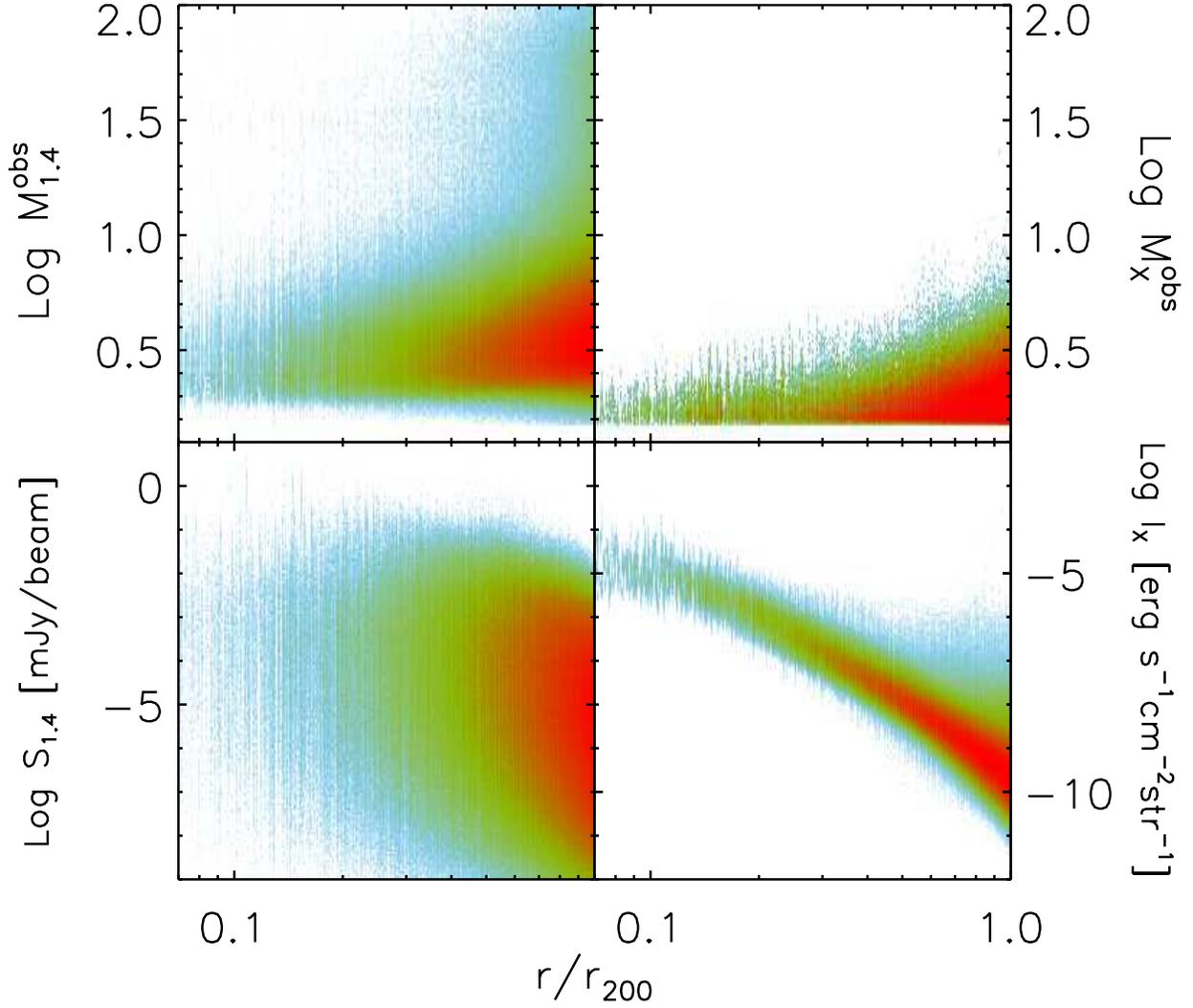}
\caption{Radial distributions of $\Msynctwo, \Mxtwo, \Ssync,$ and $\Ixbol$ for shock pixels within $r \leq \rtwo$ from
the cluster center in projected maps.
Colors code the relative frequency, ranging from $10^{-3}$ (blue) to 1 (red) in left panels,
from $10^{-1.5}$ (blue) to 1 (red) in the upper-right panel, and
from $10^{-3.5}$ (blue) to 1 (red) in the lower-right panel.}\label{fig:rad_all}
\end{figure}

\clearpage

\begin{figure}
\plotone{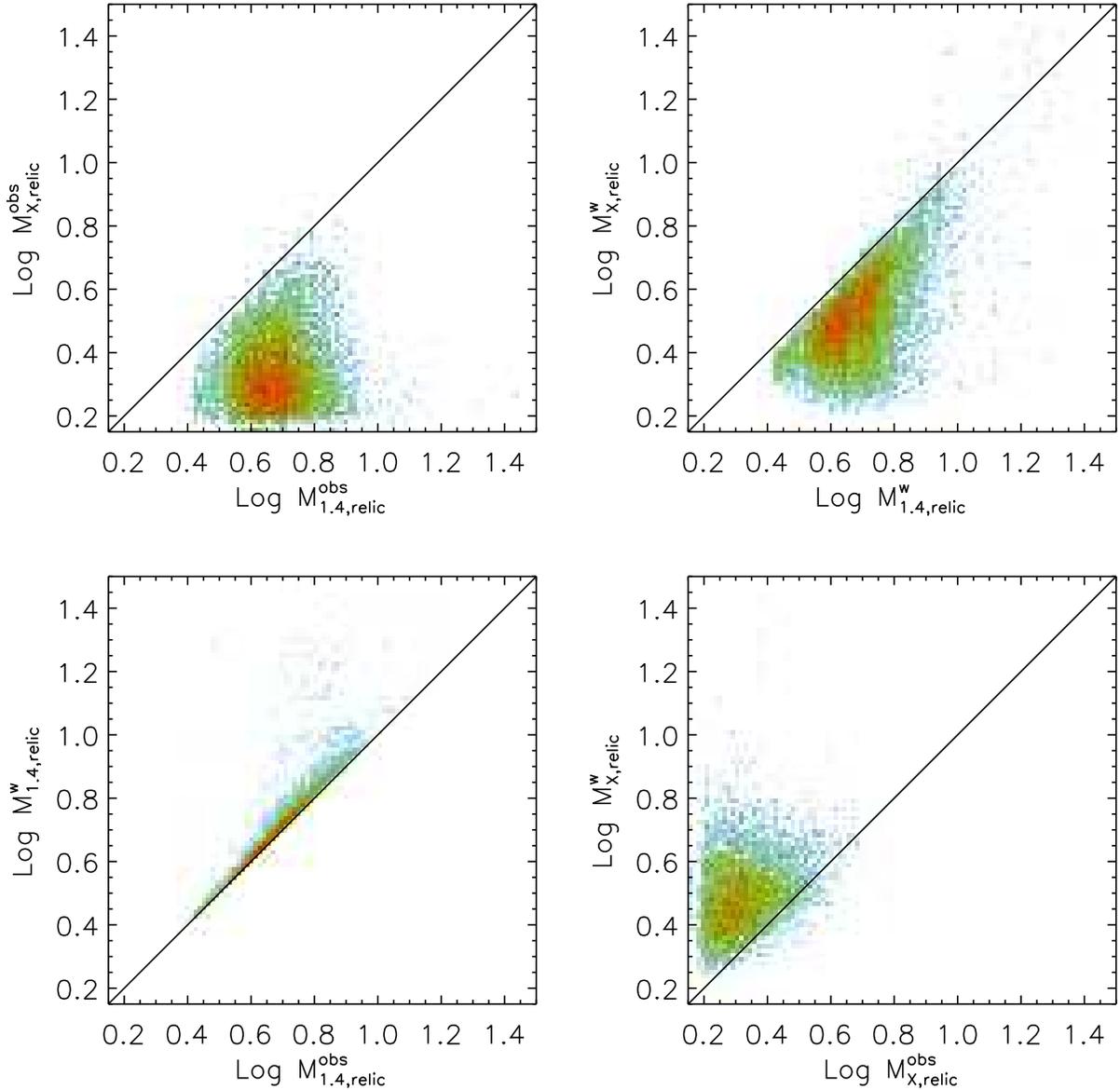}
\caption{Relations among observed Mach numbers, $\Msynctworelic$ and $\Mxtworelic$, and weighted Mach numbers, $\Msyncthreerelic$ and $\Mxthreerelic$, (see the text for definitions) for synthetic radio relics found within $r \leq \rtwo$
from the cluster center in projected maps.
Colors code the relative frequency, ranging from $10^{-2}$ (blue) to 1 (red) in the upper panels,
from $10^{-3}$ (blue) to 1 (red) in the lower-left panel,
and from $10^{-1.5}$ (blue) to 1 (red) in the lower-right panel.
Pearson's linear correlation coefficients between two derived Mach numbers at each panel are
0.65 (upper-left), 0.64 (upper-right), 0.99 (lower-left), and 0.38 (lower-right).
Black lines show the perfection correlation.}\label{fig:machs_relic}
\end{figure}

\clearpage

\begin{figure}
\plotone{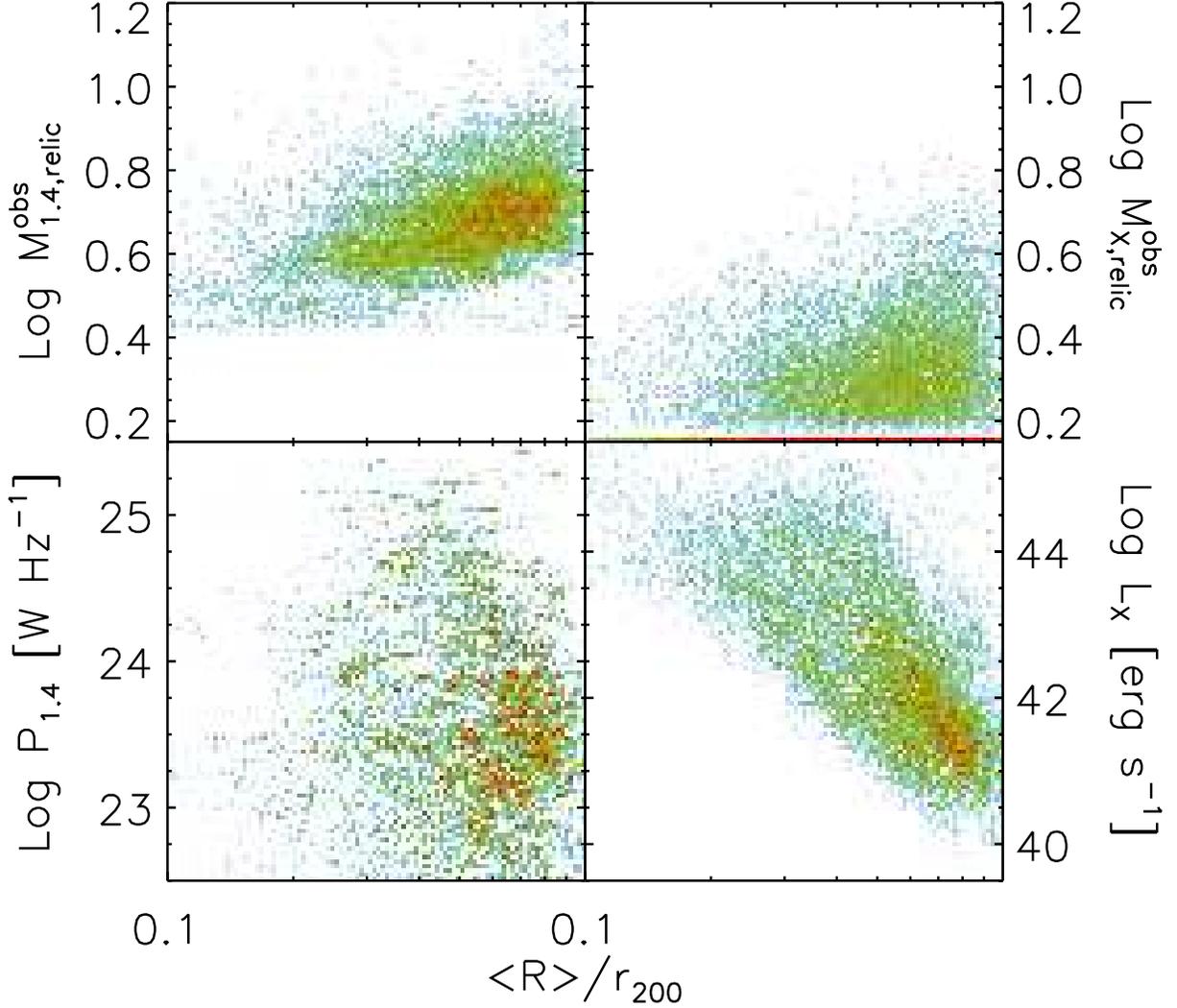}
\caption{Radial distributions of $\Msynctworelic$, $\Mxtworelic$, synchrotron power $\Psync$, and
bolometric X-ray luminosity $\Lxbol$ for synthetic radio relics found within $r \leq \rtwo$
from the cluster center in projected maps.
Colors code the relative frequency, ranging from from $10^{-2}$ (blue) to 1 (red) in the upper panels,
from $10^{-1}$ (blue) to 1 (red) in the lower-left panel,
and from $10^{-1.5}$ (blue) to 1 (red) in the lower-right panel.}\label{fig:rad_relic}
\end{figure}

\clearpage

\begin{figure}
\plotone{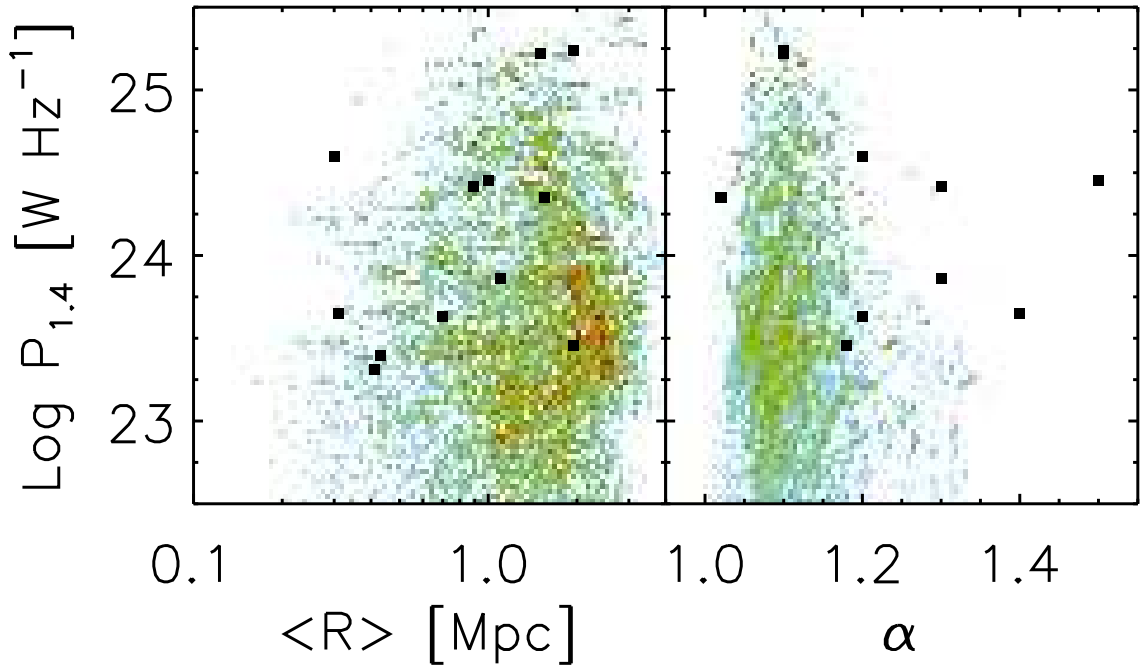}
\caption{Synchrotron power $\Psync$ vs synchrotron-weighted average distance $\rrelic$ and
integrated spectral index $\alpha$ for synthetic radio relics found within $r \leq \rtwo$
from the cluster center in projected maps.
Filled squares are data points for the observed radio relics listed in Table 1.
Colors code the relative frequency, ranging from $10^{-1.5}$ (blue) to 1 (red) in the left panel and
from $10^{-2}$ (blue) to 1 (red) in the right panel.}\label{fig:relic}\end{figure}

\end{document}